\documentclass[final,12pt]{elsarticle}

\usepackage{mathtools}                                                                    
\usepackage{amsmath}
\usepackage{amsfonts}
\usepackage{amssymb}
\usepackage{graphicx}
\usepackage{color}
\usepackage{braket}
\usepackage{hyperref}
\usepackage{shortvrb}
\usepackage{cprotect}
\usepackage{overpic}
\usepackage[capitalise]{cleveref}
\usepackage{adjustbox}
\newcommand{\B}{\mathbf{B}}
\newcommand{\E}{E}

\renewcommand{\u}{\mathbf{u}}

\newcommand{\n}{\mathbf{n}}

\newcommand{\FF}{\mathcal{F}}
\newcommand{\GG}{\mathcal{G}}
\newcommand{\R}{\mathbb{R}}
\renewcommand{\Pr}{\ensuremath{\mathrm{Pr}}}
\newcommand{\Pm}{\ensuremath{\mathrm{Pm}}}
\newcommand{\Ra}{\ensuremath{\mathrm{Ra}}}
\newcommand{\eps}[1]{\ensuremath{\varepsilon(#1)}}
\newcommand\curl{\operatorname{curl}}
\newcommand{\scurl}{\curl}
\DeclareMathOperator{\vcurl}{\mathbf{curl}}
\renewcommand\div{\nabla \cdot}
\newcommand{\del}{\partial}

\journal{Physica D: Nonlinear Phenomena}

\begin{document}

\begin{frontmatter}

    \title{Bifurcation analysis of a two-dimensional magnetic Rayleigh--B\'enard problem}

    \author[inst1]{Fabian Laakmann}
    \ead{fabian.laakmann@maths.ox.ac.uk}
    \affiliation[inst1]{organization={Mathematical Institute},
        addressline={University of Oxford},
        city={Oxford},
        postcode={OX2 6GG},
        country={United Kingdom}}

    \author[inst2]{Nicolas Boull\'e\corref{cor1}}
    \ead{nb690@cam.ac.uk}
    \cortext[cor1]{Corresponding author.}
    \affiliation[inst2]{organization={Department of Applied Mathematics and Theoretical Physics},
        addressline={University of Cambridge},
        city={Cambridge},
        postcode={CB3 0WA},
        country={United Kingdom}}

    \begin{abstract}
        We perform a bifurcation analysis of a two-dimensional magnetic Rayleigh--B\'enard problem using a numerical technique called deflated continuation. Our aim is to study the influence of the magnetic field on the bifurcation diagram as the Chandrasekhar number $Q$ increases and compare it to the standard (non-magnetic) Rayleigh--B\'enard problem. We compute steady states at a high Chandrasekhar number of $Q=10^3$ over a range of the Rayleigh number $0\leq \Ra\leq 10^5$. These solutions are obtained by combining deflation with a continuation of steady states at low Chandrasekhar number, which allows us to explore the influence of the strength of the magnetic field as $Q$ increases from low coupling, where the magnetic effect is almost negligible, to strong coupling at $Q=10^3$. We discover a large profusion of states with rich dynamics and observe a complex bifurcation structure with several pitchfork, Hopf, and saddle-node bifurcations. Our numerical simulations show that the onset of bifurcations in the problem is delayed when $Q$ increases, while solutions with fluid velocity patterns aligning with the background vertical magnetic field are privileged. Additionally, we report a branch of states that stabilizes at high magnetic coupling, suggesting that one may take advantage of the magnetic field to discriminate solutions.
    \end{abstract}

    \begin{keyword}
        Numerical bifurcation analysis \sep Magnetic Raleigh--B\'enard problem \sep Deflation
    \end{keyword}

\end{frontmatter}

\section{Introduction}

Rayleigh--B\'enard problems model buoyancy-driven convection phenomena of a fluid heated from below~\cite{rayleigh1916,benard1900,benard1927}. These systems generally support a high number of steady states and undergo a sequence of complex bifurcations as the Rayleigh number, $\textrm{Ra}$, increases~\cite{chandrasekhar1961hydrodynamic}. A large number of numerical investigations have been conducted over the past decades to characterize the influence of the Rayleigh number on the profile of solutions and type of bifurcations~\cite{crosshohenberg93,bodenschatzetal00,ouertatani2008numerical,
    zienicke1998bifurcations,paul2012bifurcation,mishra2010patterns,ma2006multiplicity,
    boronska2010extreme,boronska2010extreme2,puigjaner2004stability,puigjaner2006bifurcation}. Standard numerical techniques for performing bifurcation analysis consist of using arclength continuation~\cite{Peterson2008,keller1977numerical,doedel1981auto,uecker2014pde2path} or performing time-dependent simulations~\cite{dijkstra2014numerical,tuckerman2000bifurcation,mamun1995asymmetry}. A different approach based on deflation~\cite{Deflation2015} was recently proposed and allows the automatic computation of many steady states to the Rayleigh--B\'enard problem~\cite{Boulle2022}.

In this paper, we focus on a two-dimensional (2D) magnetic Rayleigh--B\'enard problem, in which an external magnetic field is applied to an electrically conducting fluid, and study how the strength of the magnetic field influences the bifurcation patterns. These so-called magnetohydrodynamics (MHD) convection problems lead to interesting instability~\cite{weiss_proctor_2014, Busse1982Stability, PhysRevLett.52.1774}, heat transport~\cite{aurnou_olson_2001,burr_muller_2002}, flow reversal~\cite{yanagisawa2011spontaneous} phenomena, and have many applications in astrosciences and geosciences~\cite{Proctor_1982,glatzmaier1999role, Cattaneo_2003, rucklidge2006mean}. Then, the bifurcation analysis of the magnetic Rayleigh--B\'enard problem using either numerical simulations or laboratory experiments is an active field of research~\cite{Yang2021, HAN2018370, NAFFOUTI2014714, Akhmedagaev2020}. As an example, Akhmedagaev \emph{et al.} performed numerical simulations of turbulent Rayleigh--B\'enard convection in a cylinder under a vertical magnetic field at high Rayleigh and Hartmann numbers~\cite{Akhmedagaev2020}. More recently, Yang \emph{et al.} performed a laboratory experiment to study the hydromagnetic convection of liquid metal in a cuboid vessel and analyze the transition from steady to oscillating convection rolls in the regime $\textrm{Ra}\in[2.3\times10^4,2.6\times10^5]$~\cite{Yang2021}. Finally, Naffouti \emph{et al.} investigated the effect of the direction of the magnetic field on the convection flow patterns~\cite{NAFFOUTI2014714}.

The effect of the magnetic field can be quantified using the so-called Chandrasekhar number, $Q$, which is a dimensionless quantity characterizing the strength of the Lorentz force. The influence of $Q$ on the occurring instabilities is a well-studied problem in the literature~\cite{burr_muller_2002, Tasaka2016, Cioni2000, Nandukumar_2015, DAS2019, Basak2014}. Hence, Burr \& M{\"u}ller performed a physical experiment of a heated liquid metal layer at  Rayleigh number between $10^3$ and $10^5$ and Chandrasekhar number lower than $1.44 \times 10^6$~\cite{burr_muller_2002}, while Tasaka \emph{et al.} explored convection regime in the parameter range $5 \times 10^3 < \Ra < 3\times 10^5$ and $2 \times 10^3 < Q < 10^4$~\cite{Tasaka2016}. Moreover, Cioni \emph{et al.} performed experiments up to $Q = 4 \times 10^6$ to characterize turbulent magnetohydrodynamics regimes beyond the convection threshold~\cite{Cioni2000}. Numerical simulations of magnetic Raleigh-B\'enard convection have also been conducted in the regime $0 \leq Q \leq 2.5 \times 10^4$~\cite{DAS2019, Nandukumar_2015}. Interestingly, a large number of experimental, numerical, and theoretical studies suggest that the magnetic field can be used to delay primary, secondary, and higher-order instabilities~\cite{chandrasekhar1961hydrodynamic, Busse1982Stability,Nakagawa1957Experiments,
    Nakagawa1959Experiments,knobloch_weiss_costa_1981,clever1989nonlinear,Basak2014}.

Here, we investigate numerically the effect of the magnetic field on the bifurcation diagram for the Rayleigh number $\Ra\in[0,10^5]$ as the Chandrasekhar number increases from $1$ to $10^3$ with Prandtl and magnetic Prandtl numbers fixed to $1$. By combining deflation~\cite{Deflation2015} and a continuation technique in the Chandrasekhar number, we are able to relate primary instabilities and secondary branches at $Q=10^3$ to the corresponding Rayleigh--B\'enard branches in the low magnetic field limit near $Q=0$. This allows us to perform qualitative and quantitative comparisons between the convection patterns discovered by deflation at $Q=10^3$ and the related branches analyzed in~\cite{Boulle2022}, as well as study the effect of the magnetic field strength on the stability of the steady states.

The paper is organized as follows. We first introduce the non-dimensional formulation of the Rayleigh--B\'enard problem and describe the numerical methods in \cref{sec:BifurcationAnalysis}. Then, we present the numerical results and analyze the effect of the strength of the magnetic field on the bifurcation diagram in \cref{sec:results}. Finally, in \cref{sec_conclusions}, we summarize the conclusions of this work and propose potential extensions.

\section{Magnetic Rayleigh--B\'enard problem in 2D}\label{sec:BifurcationAnalysis}

\subsection{Formulation and discretisation}\label{sec:AnIsoFormulationAndDiscretisation}

We consider a two-dimensional magnetic Rayleigh--B\'enard problem modeling a conducting fluid heated from below in a unit square cell domain $\Omega=(0,1)^2$, with spatial coordinates $(x, z)$. While most of the earlier theoretical and numerical works on this topic consider stress-free boundary conditions for the fluid velocity to perform a decomposition of the periodic solutions in the Fourier basis or pseudo-spectral numerical simulations~\cite{weiss_proctor_2014, DAS2019, Nandukumar_2015}, we assume no-slip boundary conditions for the fluid velocity and a background vertical magnetic field pointing upwards. This choice allows us to compare the profile of solutions discovered by deflation with~\cite{Boulle2022}. Furthermore, we choose the horizontal walls to be thermally conducting and the vertical walls to be insulating.

We employ the Boussinesq approximation \cite{boussinesq1903theorie, Oberbeck1879} to model magnetohydrodynamics convection and assume that the flow is buoyancy-driven and that density differences only appear in the buoyancy term, while other parameters do not depend on density or temperature. Then, one can write down the non-dimensional formulation of the two-dimensional anisothermal MHD equations with Boussinesq approximation on the domain $\Omega$ as
\begin{subequations}
    \label{eq:MHDBou}
    \begin{align}
        \partial_t \u - 2 \Pr \div \eps \u +  \u \cdot \nabla \u + \nabla p & \nonumber                        \\+ \Pr\, Q\, \B \times  (\E + \u \times \B) &=
        \Ra\, \Pr\, T \hat{\mathbf{z}}, \label{eq:MHDBou1}                                                     \\
        \div \u                                                             & =0, \label{eq:MHDBou2}           \\
        \E + \u \times \B - \frac{\Pr}{\Pm} \scurl \B                       & = 0, \label{eq:MHDBou3}          \\
        \partial_t \B + \vcurl \E                                           & = \mathbf{0}, \label{eq:MHDBou4} \\
        \partial_t T - \nabla^2 T + \u \cdot \nabla T                       & = 0,                             \\
        \div \B                                                             & = 0, \label{eq:MHDBou5}
    \end{align}
\end{subequations}
where $\u=(u,w)$ denotes the fluid velocity field, $p$ the fluid pressure, $\B=(B_x,B_z)$ the magnetic field, $\E$ the electric field, $T$ the temperature, $\hat{\mathbf{z}}= \begin{pmatrix}
        0 & 1
    \end{pmatrix}^\top$
the buoyancy direction, and $\eps \u = \frac{1}{2} (\nabla \u + \nabla \u^\top)$ the symmetric gradient. For clarity, we denote vector fields with bold letters in \cref{eq:MHDBou}. We emphasize that in two dimensions the electric field is a scalar field and is hence denoted as $E$. Moreover, there exist two curl operators and cross products which are interpreted differently depending on whether the arguments are scalar- or vector-valued as
\begin{align*}
    \scurl \B    & = \del_x B_z - \del_z B_x, &  &  & \vcurl E    & =
    \begin{pmatrix}
        \del_z E \\
        -\del_x E
    \end{pmatrix},                                                    \\
    \u \times \B & =u B_z - w B_x,            &  &  & \B \times E & =
    \begin{pmatrix}
        B_z E \\
        -B_x E
    \end{pmatrix}.
\end{align*}
For this reason, we employ the boundary condition $\E = 0$ on $\partial \Omega$, which is the two-dimensional version of the standard condition $\mathbf{E} \times \n = \mathbf{0} $ on $\partial \Omega$. We then complement \cref{eq:MHDBou} with the boundary conditions $\u = \mathbf{0}$, $\B\cdot \hat{\mathbf{n}} = \hat{\mathbf{z}}$, and $E=0$ on the boundary of the domain $\partial\Omega$, where $\hat{\mathbf{n}}$ denotes the unit outer normal vector. In addition, we enforce $T=1$ at $z=0$ and $T=0$ at $z=1$, as well as $\partial_x T=0$ at $x=0$ and $x=1$.

The dimensionless numbers in \cref{eq:MHDBou}, \emph{i.e.} the Prandtl, magnetic Prandtl, Rayleigh, and Chandrasekhar numbers, are respectively defined as $\Pr=\nu\alpha^{-1}$, $\Pm=\nu\eta^{-1}$, $\Ra = \beta g \overline{T} L^3(\nu \alpha)^{-1}$, $Q = \overline{B}^2 L^2(\rho_0 \mu_0 \nu \eta)^{-1}$. Here, $\rho_0$ denotes the reference density, $\nu$ the kinematic viscosity, $\beta$ is the thermal expansion, $\eta$ the magnetic resistivity, $g$ the gravitational acceleration, $\alpha$ the thermal diffusivity, $\mu_0$ the magnetic permeability of free space, $\overline{B}$ a characteristic value for the magnetic field, $L$ a characteristic length scale, and $\overline{T}$ the difference between two reference temperatures (\emph{e.g.}, the reference temperatures may be defined as the temperature of the hot and cold plates at the top and bottom of our domain). In the numerical experiments performed in \cref{sec:results}, we will study the influence of the magnetic field on the bifurcation diagram with respect to the Rayleigh number by fixing $\Pr=1$, $\Pm=1$, and varying $\Ra$ and $Q$ in the range $\Ra\in[0,10^5]$ and $Q\in[1,10^3]$.

The trivial stationary solution (conducting state) to \cref{eq:MHDBou} with our choice of perfectly conducting boundary conditions is given by $\phi_0$, where
\begin{equation} \label{eq:trivialsol}
    \begin{gathered}
        \u_0 = \mathbf{0}, \quad p_0 = \Ra\,\Pr \left(z - \frac{z^2}{2} - \frac{1}{3}\right), \quad T_0 = 1-z, \\
        \B_0 = \hat{\mathbf{z}},\quad \text{and}\quad E_0 = 0,
    \end{gathered}
\end{equation}
and $\phi_0=(\u_0,p_0,T_0,\B_0,E_0)$. The boundary conditions for $\B$ are prescribed by an external magnetic field pointing in $ \hat{\mathbf{z}}$ direction.
Additionally, we remark that the problem has two symmetries which determine the behavior of the arising bifurcations with respect to $\Ra$ and $Q$: the mirror symmetry with respect to $x=1/2$,
\[[u, w, T, B_x, B_z](x, z) \to [-u, w, T, B_x, -B_z] (1-x, z)\]
and the Boussinesq symmetry,
\[[u, w, T, B_x, B_z](x, z) \to [u, -w, 1-T, -B_x, B_z] (x, 1-z).\]
These symmetries are similar to the Rayleigh--B\'enard problem (cf.~\cite[Sec.~2]{Boulle2022}) but also transform the electric and magnetic fields.

We reformulate the stationary version of \cref{eq:MHDBou} by combining \cref{eq:MHDBou4,eq:MHDBou5} to obtain the following equivalent augmented Lagrangian formulation:
\begin{equation}\label{eq:ALform}
    -\frac{\Pr}{\Pm} \nabla(\div \B) + \vcurl \E = \mathbf{0}.
\end{equation}

It is obvious that \cref{eq:MHDBou4,eq:MHDBou5} imply \cref{eq:ALform} in the stationary case. For the reverse direction, \cref{eq:ALform} can be multiplied with $\vcurl E$ and $\nabla \cdot \B$ and be integrated over the domain $\Omega$. After performing integration by parts one obtains \cref{eq:MHDBou4,eq:MHDBou5} by the choice of boundary conditions and the fact the curl of a gradient and the divergence of a curl vanishes.

Hu \emph{et al.} showed that the augmented Lagrangian formulation makes the stationary version of \cref{eq:MHDBou} well-posed~\cite{Hu2020} and suitable for a finite-element discretization. This is also the reason why the electric field $E$ is kept in our formulation and not eliminated via \eqref{eq:MHDBou3}. Moreover, the presented formulation allows for a structure-preserving finite-element discretization that enforces both $\nabla \cdot \B = 0$ and $\vcurl E=0$ exactly at the discrete level \cite{Hu2020}. Indeed, preserving the discrete solenoidality of $\B$ is crucial for numerical simulations to produce physically accurate results~\cite{Brackbill1980}.

We employ a finite element method to solve the stationary magnetic Rayleigh--B\'enard problem in two dimensions. The velocity and pressure are discretized using the standard Taylor--Hood elements~\cite{taylor1973numerical}, \emph{i.e.}, piecewise vector valued quadratic polynomials for $\u$ and piecewise linear polynomials for $p$. Then, the temperature and electric fields are respectively approximated by piecewise linear and quadratic polynomials. Finally, we use second-order Raviart--Thomas elements~\cite{Raviart1977} for the discretization of the magnetic field, which is a standard choice for a $\mathcal{H}(\mathrm{div})$-conforming finite element that allows for the aforementioned structure-preserving discretization. The domain $\Omega$ is discretized using a crossed triangular mesh with $50\times 50$ square cells, where each square cell is divided into four triangles following the diagonals of the squares to preserve the symmetries of the problem. While this is a rather coarse mesh, we emphasize that computing the full bifurcation diagram might require performing hundreds of thousands of Newton iterations, which is computationally expensive. However, we checked that refining the mesh does not influence the profile of the steady states we are computing in the range of parameters, $0\leq \Ra\leq 10^5$ and $1\leq Q\leq 10^3$, considered. Moreover, this mesh allows us to employ a sparse LU direct solver to solve the underlying linear systems. We produce our numerical results using the Firedrake finite element software~\cite{Firedrake}, which relies on the PETSc package \cite{balay2019} for solving the discretized equations.

\begin{figure*}[htbp]
    \centering
    \begin{overpic}[width=\textwidth]{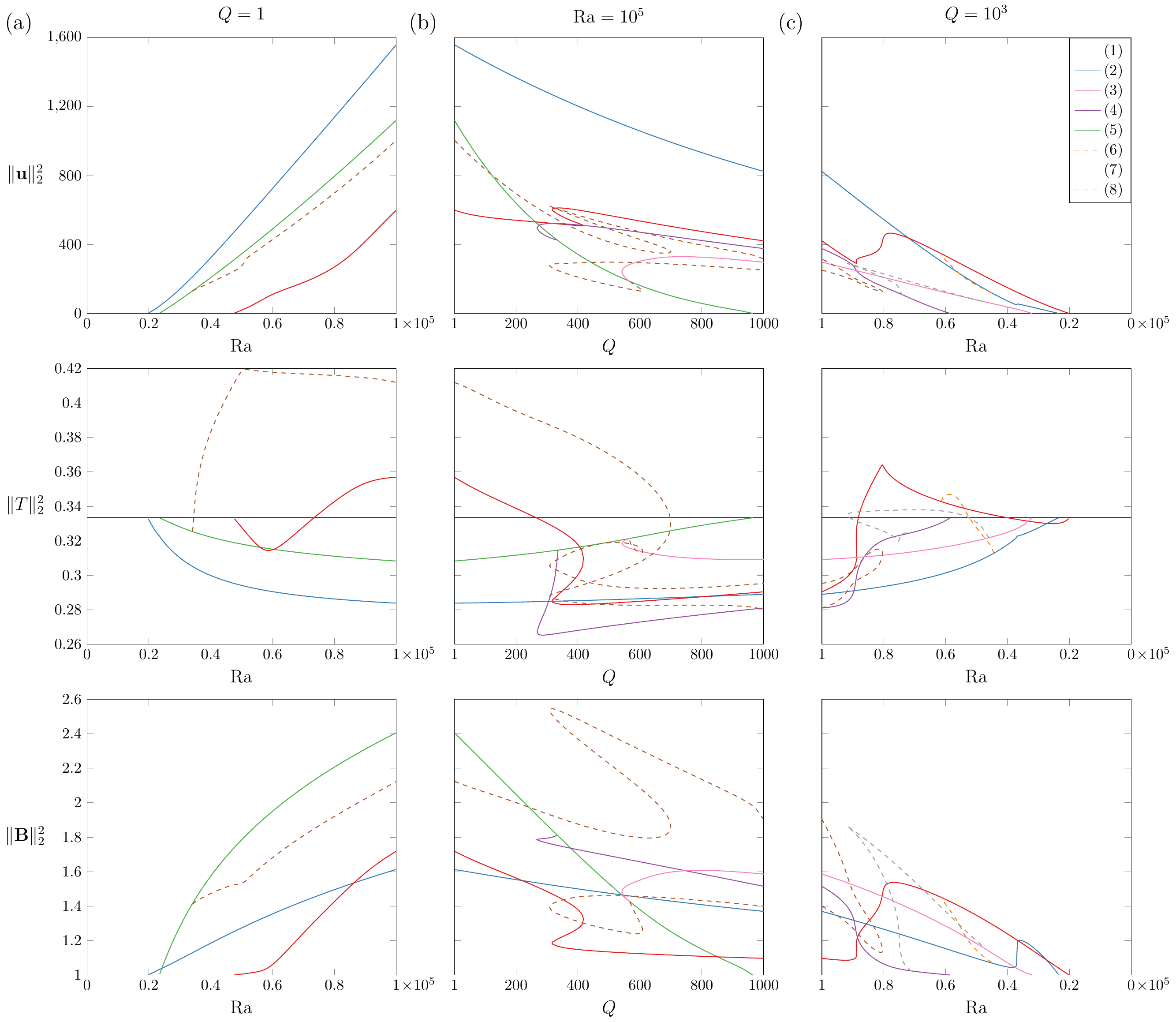}
    \end{overpic}
    \caption{Bifurcation diagrams of the magnetic Rayleigh--B\'enard problem using the kinetic energy $\|\u\|_2^2$ (top row), potential energy $\|T\|_2^2$ (middle row), and magnetic energy $\|\B\|_2^2$ (bottom row) as diagnostic. The diagrams display the evolution of the branches studied in this paper, \emph{i.e.}, the first four primary instabilities at $Q=10^3$ and their secondary bifurcations, in the range $0\leq \Ra\leq 10^5$ at $Q=1$ (a), $1\leq Q\leq 10^3$ at $\Ra=10^5$ (b), and $0\leq \Ra\leq 10^5$ at $Q=10^3$ (c). The first four primary branches for the bifurcation diagrams in (c) are displayed with solid lines while the secondary branches are indicated with dashed lines. The dark line in the diagrams using the potential energy corresponds to the conducting state satisfying $\|T_0\|_2^2=1/3$. The $x$-axes of the diagrams in panel (c) are reversed to analyze the evolution of the branches as the strength of the magnetic field increases.}
    \label{fig:bifurcaton_diagram}
\end{figure*}

\subsection{Bifurcation technique} \label{sec_bifurcation}

Steady states to the magnetic Rayleigh--B\'enard system \eqref{eq:MHDBou} are computed using a numerical technique called \emph{deflation}~\cite{Deflation2015}. This algorithm aims to discover multiple solutions to nonlinear partial differential equations (PDEs) in an iterative manner by modifying Newton's method in order to avoid its convergence to previously obtained solutions. While deflation has been successfully applied to a wide range of physical systems such as liquid crystals~\cite{robinson2017molecular,emerson2018computing}, quantum mechanics~\cite{charalampidis2018computing,charalampidis2020bifurcation,boulle2020deflation,boulle2022two}, and fluid dynamics~\cite{Boulle2022}, to our knowledge this work is its first adaptation to MHD-type problems. To introduce this numerical method, we first rewrite our bifurcation problem as $F(\Phi,\lambda)=0$, where $\Phi = (\u, p, T, \B, E)$ is the solution, $\lambda \in \R$ denotes the bifurcation parameter, and $F$ is the partial differential operator associated with the stationary version of the MHD system \eqref{eq:MHDBou}. In this work, we employ either the Rayleigh number or the Chandrasekhar number as bifurcation parameters and keep the other dimensionless numbers fixed. For a given $\lambda\in \R$, assuming that Newton's method has converged to a solution $\Phi_1$ satisfying $F(\Phi_1, \lambda)=0$, then deflation constructs a new (deflated) problem of the form
\begin{equation} \label{eq_deflation}
    F_1(\Phi, \lambda) \coloneqq \mathcal{M}(\Phi, \Phi_1) F(\Phi, \lambda).
\end{equation}
The deflation operator $\mathcal{M}$ is chosen to penalize solutions that are close to $\Phi_1$ such that $\lim_{\phi\to\phi_1}\mathcal{M}(\Phi, \Phi_1)=\infty$. In this work, we employ the following deflation operator:
\[	\mathcal{M}(\Phi, \Phi_1) \coloneqq \left( \frac{1}{\mathcal{N}(\Phi, \Phi_1)} + 1 \right),\]
where the operator $\mathcal{N}$ is defined as
\[	\mathcal{N}(\Phi, \Phi_1) \coloneqq \|\u - \u_1 \|^2_{H^1}  + \|T - T_1\|^2_2 + \|\B - \B_1\|^2_2,\]
such that $\mathcal{N}(\phi,\phi_1)$ converges to zero as $\phi$ approaches $\phi_1$. Here, $\|\cdot\|_2$ denotes the $L^2$-norm and $\|\u\|^2_{H^1} = \|\u\|^2_2 + \|\nabla \u\|^2_2$ is the squared $H^1$-norm. Then, one can apply Newton's method on the deflated problem $F_1$ from the same initial guess used to discover $\phi_1$ without converging to the same solution, and hopefully discover a different solution to $F(\phi,\lambda)=0$. A key benefit of deflation is that solving \cref{eq_deflation} via Newton's method does not require more computational work than solving the original system since one can express the Newton update of the discrete deflated system as a scaling of the update for the original problem~\cite{Deflation2015}. This process can be repeated to discover several steady states to the magnetic Rayleigh--B\'enard problem by iteratively composing deflation operators to deflate multiple roots.

Once deflation has been applied at a fixed value of the bifurcation parameter $\lambda$, we interlace it with a continuation procedure to reconstruct a bifurcation diagram over a range of parameters $\lambda\in[\lambda_{\min},\lambda_{\max}]$. We then use the discovered solutions at $\lambda$ as initial guesses for a parameter continuation from $\lambda$ to $\lambda \pm \Delta \lambda$, where $\Delta\lambda$ is the step-size and the sign determines the sense (forward or backward) of the continuation. The algorithm then iteratively continues by applying a deflation to each discovered solution at $\lambda \pm \Delta \lambda$ until the final values of $\lambda$ are reached. Here, one could mitigate the computational cost of the deflation procedure by using a coarse deflation step size and performing a deflation step after a certain number of fine continuation steps to discover new branches.

Since one initially employs Newton's method on the original problem, the performance of deflation at finding new solutions depends heavily on the available initial guesses for the first deflation step. To address this issue, one technique applied for the Rayleigh--B\'enard problem~\cite{Boulle2022} is to construct initial guesses for backward continuation starting from $\Ra=10^5$ by summing the trivial solution and its normalized unstable eigenmodes. This ensures that the initial guesses are close to the solutions in the branches emerging from the conducting state close to the linear regime. While the same approach works for the magnetic Rayleigh--B\'enard problem at the small Chandrasekhar number $Q=1$, it fails to provide useful initial guesses for Newton's method when applied in a regime of stronger magnetic fields, such as $Q=10^3$, as the solutions differ significantly from the ones of the linear regime. Therefore, our remedy consists of first performing deflation with respect to the Rayleigh number in the range $0\leq\Ra\leq 10^5$ at $Q=1$, then prolonging this bifurcation diagram at $\Ra=10^5$ over $1\leq Q\leq 10^3$ with deflation. This procedure finally provides good initial guesses at $\Ra=10^5$ and $Q=10^3$ for backward deflated continuation over $\Ra\in[0,10^5]$. This results in three bifurcation diagrams: over $\Ra$ at $Q=1$ ($\Delta\Ra=10^3/3$), over $Q$ at $\Ra=10^5$ ($\Delta Q=10/3$), over $\Ra$ at $Q=10^3$ ($\Delta\Ra=500$), which are respectively displayed in \cref{fig:bifurcaton_diagram}(a-c), and enables the study of the influence of the magnetic field on the bifurcation diagrams by relating the states at Chandrasekhar number $Q=10^3$ to the ones at $Q=1$.

In this work, we are also interested in the evolution of the velocity, temperature, and magnetic profiles of the solutions and therefore used the squared $L^2$-norm of these fields (resp.~kinetic energy $\|\u\|_2^2$, potential energy $\|T\|_2^2$, and magnetic energy $\|\B\|_2^2$) as diagnostic for the bifurcation diagrams. As an example, the blue curve in the top row of \cref{fig:bifurcaton_diagram} reports the evolution of the kinetic energy of branch (2) as the Rayleigh number first increases in $[0,10^5]$ with $Q=1$ (a), then as $Q$ decreased in the interval $[10^3,1]$ with $\Ra=10^5$ (b), and as $\Ra$ decreases from $10^5$ to $0$ with $Q=10^3$. A solid curve corresponds to a primary branch emerging from the conducting state while a dashed curve indicates a secondary branch. For readability, the color of the branches matches the figures in \cref{sec:results}, which analyze the solutions and their stability in each branch. Finally, we note that while a large profusion of states has been discovered by deflation, we focus on a selection arising from primary bifurcations of the trivial state and their second bifurcations to keep their analysis tractable.

In principle, there would not be any obstacle in extending this deflated continuation approach in a multiparameter setting. However, the computational cost would increase significantly, as one would need to perform a continuation in both parameters $\Ra$ and $Q$ simultaneously, which would square the number of deflation/continuation steps. Moreover, the analysis would be relatively challenging as one would need to consider the interaction between the two parameters while following the different branches.

\subsection{Choice of the bifurcation parameters}\label{sec:choice_of_parameters}

\begin{figure*}[htbp]
    \centering
    \begin{overpic}[width=\textwidth]{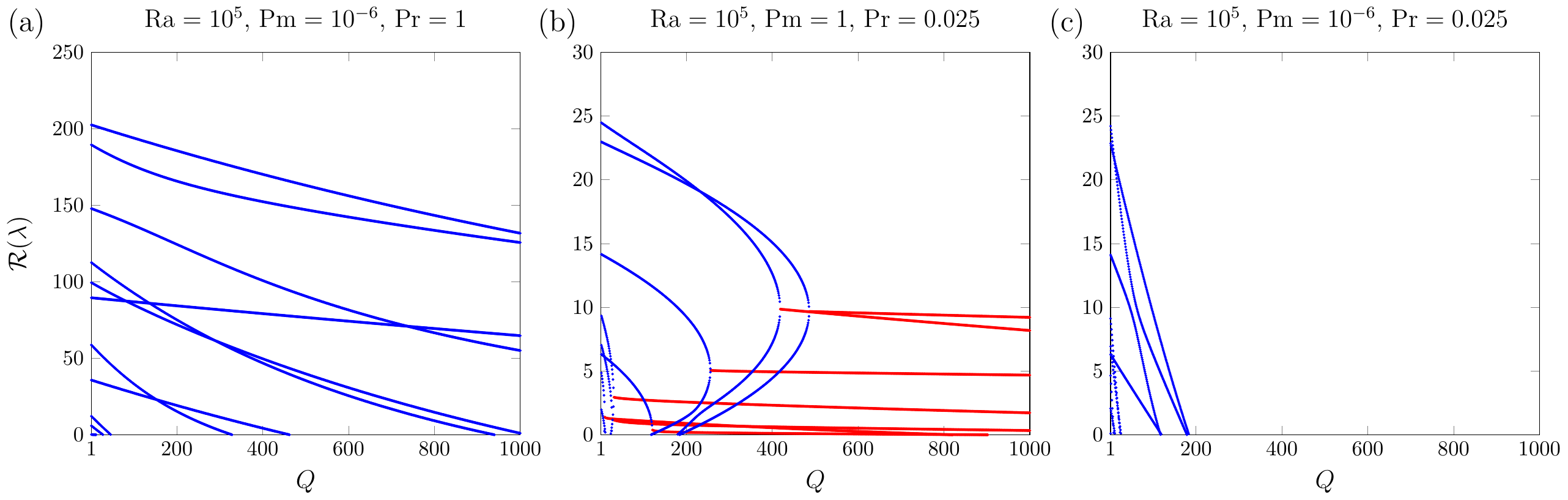}
    \end{overpic}
    \caption{Growth rates of the conducting state in different parameter regimes as a function of the Chandrasekhar number $Q$. The growth rates are computed using the leading eigenvalues of the linearized system around the conducting state (see~\cref{app:unstableEigenmodes}). In panel (b), the red lines indicate that the imaginary part of the eigenvalues is non-zero and grows as $Q$ increases, eventually leading to Hopf bifurcations when the corresponding growth rates vanish.}
    \label{fig:bifurcation_parameters}
\end{figure*}

The parameters $\textrm{Pr}$ and $\textrm{Pm}$ have been chosen so that the underlying system has a rich behavior with complex structures and bifurcations at the boundary between steady and oscillatory to demonstrate the effectiveness of our numerical method and may not be representative of a concrete physical system. Hence, liquid metal experiments suggest $\textrm{Pr}\approx 0.02$ and $\textrm{Pm}\approx 10^{-6}$~\cite{burr_muller_2002}. We performed numerical tests to explore the system's behavior at more realistic values of $\textrm{Pm}=10^{-6}$ and $\textrm{Pr}=0.025$. First, decreasing the value of $\textrm{Pm}$ from $1$ to $10^{-6}$ gives a minor influence on the birth of primary branches to the conducting state. As an example, consider \cref{fig:bifurcation_parameters}(a), which is an analogue of \cref{fig:eigsplots}(b) with $\textrm{Pm}=10^{-6}$. The locations at which the bifurcations emerge, as well as the resulting growth rates, only differ slightly between the two settings. However, performing parameter continuation along with deflation to perform a systematic bifurcation analysis at this parameter value is computationally challenging as one requires a much finer grid to resolve the solutions. Changing $\textrm{Pr}$ to 0.025 has a much bigger effect on the bifurcation pattern as none of the instabilities persist past $Q\approx 500$. Similarly, at the realistic parameter values of $\textrm{Pm}=10^{-6}$ and $\textrm{Pr}=0.025$, we find that the growth rates of the conducting state decay to zero before $Q=200$, leading to no birth of primary branches past this value. Additionally, one of the main motivations of our paper is to compare the magnetic case to the nonmagnetic case (studied in~\cite{Boulle2022}), where the parameter $\textrm{Pr}=1.0$ was considered.

\subsection{Stability analysis}\label{app:unstableEigenmodes}

Once a stationary solution $\phi_1=(\u_1, p_1, T_1, \B_1, E_1)$ to \cref{eq:MHDBou} has been found by deflation, we analyze its stability using the perturbation ansatz
\begin{equation} \label{eq_pert}
    \begin{pmatrix}
        \u \vphantom{\tilde{E}} \\
        p \vphantom{\tilde{E}}  \\
        T \vphantom{\tilde{E}}  \\
        \B \vphantom{\tilde{E}} \\
        E \vphantom{\tilde{E}}
    \end{pmatrix}
    =
    \begin{pmatrix}
        \u_1 \vphantom{\tilde{E}} \\
        p_1 \vphantom{\tilde{E}}  \\
        T_1 \vphantom{\tilde{E}}  \\
        \B_1 \vphantom{\tilde{E}} \\
        E_1 \vphantom{\tilde{E}}
    \end{pmatrix}
    + \epsilon
    \begin{pmatrix}
        \tilde{\u} \\
        \tilde{p}  \\
        \tilde{T}  \\
        \tilde{\B} \\
        \tilde{E}
    \end{pmatrix}
    e^{\lambda t},
\end{equation}
where $\epsilon\ll 1$ is a small parameter and $\lambda\in \mathbb{C}$ is an eigenvalue. Upon inserting \cref{eq_pert} into \cref{eq:MHDBou} to linearize it around the solution $(\u_1, p_1, T_1, \B_1, E_1)$ and inserting the perturbation ansatz, we obtain the following generalized eigenvalue problem:
\begin{align*}
    \begin{bmatrix}
        \FF                & -\nabla & \Ra\, \Pr\, \hat{\mathbf{z}} & \GG                                    & - \Pr\, Q\,\B_1 \times  \\
        \nabla \cdot       & 0       & 0                            & 0                                      & 0                       \\
        - \nabla T_1 \cdot & 0       & \nabla^2 - \u_1 \cdot \nabla & 0                                      & 0                       \\
        0                  & 0       & 0                            & \frac{\Pr}{\Pm}\nabla (\nabla \cdot)   & - \vcurl                \\
        - \times \B_1      & 0       & 0                            & -\frac{\Pr}{\Pm}\scurl \, - \u_1 \cdot & -I \vphantom{\tilde{E}}
    \end{bmatrix}
     & \begin{bmatrix}
           \tilde{\u} \\ \tilde{p} \\  \tilde{T} \\ \tilde{\B}\\ \tilde{E}
       \end{bmatrix} \\
    = \lambda
    \begin{bmatrix}
        I & 0 & 0 & 0 & 0 \vphantom{\tilde{E}} \\
        0 & 0 & 0 & 0 & 0 \vphantom{\tilde{E}} \\
        0 & 0 & I & 0 & 0 \vphantom{\tilde{E}} \\
        0 & 0 & 0 & I & 0 \vphantom{\tilde{E}} \\
        0 & 0 & 0 & 0 & 0 \vphantom{\tilde{E}}
    \end{bmatrix}
     &
    \begin{bmatrix}
        \tilde{\u} \\ \tilde{p} \\  \tilde{T} \\ \tilde{\B}\\ \tilde{E}
    \end{bmatrix},
\end{align*}
where $I$ denotes the identity and the operators $\FF$ and $\GG$ are defined as
\begin{align*}
    \FF \tilde{\u} \coloneqq \,\, & 2 \Pr\, \nabla \cdot \varepsilon(\tilde{\u}) - \u_1 \cdot \nabla \tilde{\u} - \tilde{\u} \cdot \nabla \u_1 \\
                                  & - \Pr\, Q\, \B_1 \times (\tilde{\u} \times \B_1),                                                          \\
    \GG \tilde{\B} \coloneqq \,\, & - \Pr\, Q\, \tilde{\B} \times E_1 - \Pr\, Q\, \tilde{\B}\times (\u_1 \times \B_1)                          \\
                                  & - \Pr\, Q\, \B_1\times(\u_1 \times \tilde{\B}).
\end{align*}
We solve this generalized eigenvalue problem to uncover the eigenvalues with the largest real parts using a Krylov Schur solver~\cite{Stewart2002} that is implemented in the SLEPc library~\cite{SLEPc}.

The real and imaginary parts of the computed eigenvalues determine the stability of the corresponding eigenmode. If the real part (growth rate) of all eigenvalues is negative, the solution is stable, whereas if at least one eigenvalue has a positive real part the solution is unstable. Moreover, the type of instability depends on whether the associated imaginary part is zero or non-zero. The change of stability of a solution branch at a bifurcation parameter $\lambda$ indicates the presence of bifurcations, \emph{e.g.}, a simple bifurcation if the solution has a zero eigenvalue or a Hopf bifurcation if it has a pair of complex conjugate purely imaginary eigenvalues~\cite{hale2012dynamics}.

\begin{figure*}[htbp]
    \centering
    \begin{overpic}[width=\textwidth]{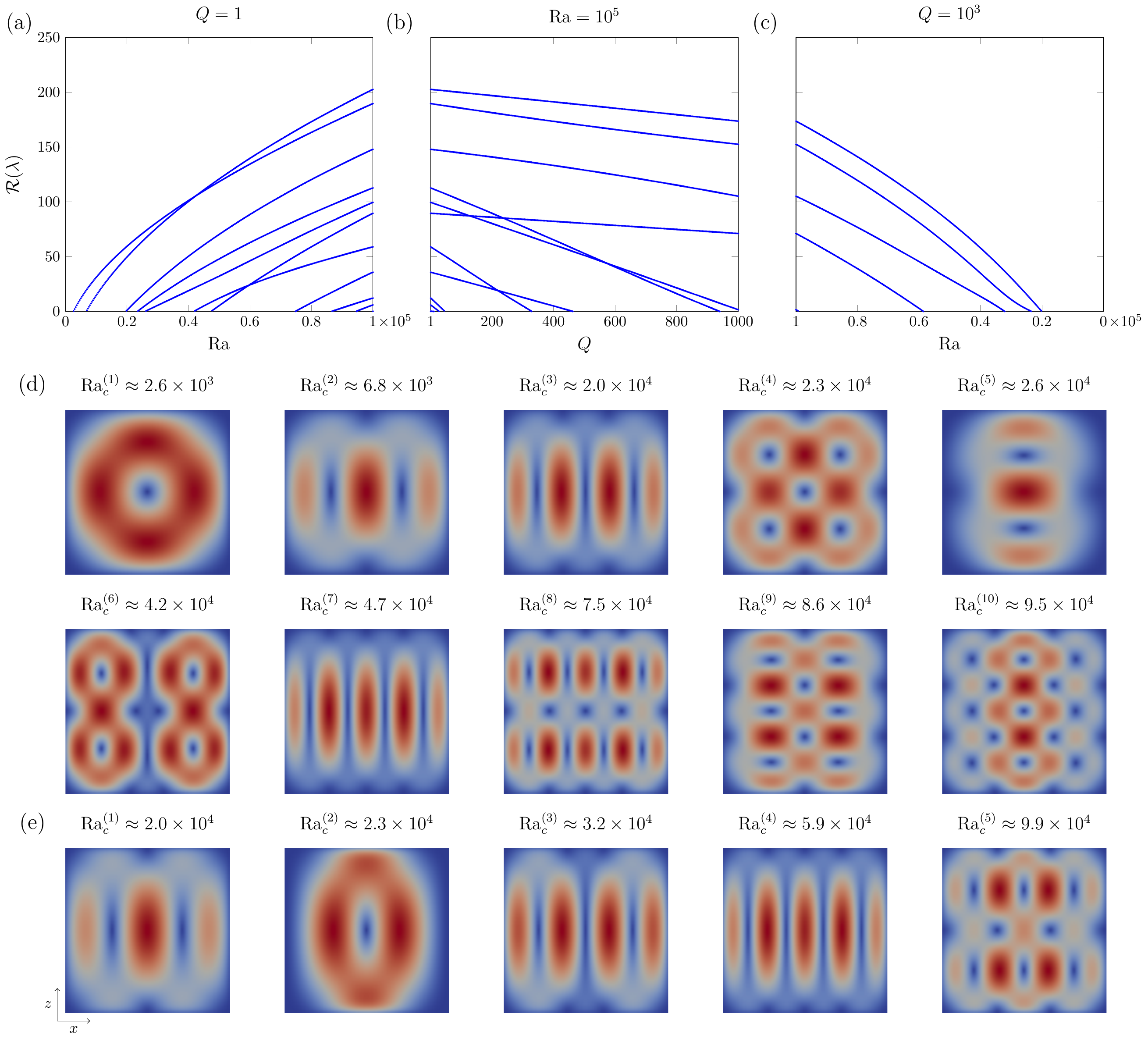}
    \end{overpic}
    \caption{(a-c) Evolution of the positive growth rates of the conducting state~\eqref{eq:trivialsol} as the Rayleigh and Chandrasekhar numbers vary in the intervals $0\leq \Ra \leq 10^5$ and $1\leq Q\leq 10^3$. (d) First ten eigenmodes of the primary bifurcations that emanate from the conducting state in the range $0\leq \Ra \leq 10^5$ at \mbox{$Q=1$}. The plots display the magnitude of the velocity, where blue colors indicate the zero-velocity magnitude and red colors correspond to a high magnitude. (e) Eigenmodes associated with the first five primary bifurcations that emanate from the conducting state in the range $0\leq \Ra \leq 10^5$ at \mbox{$Q=10^3$}. Note that, as before, the x-axis is reversed in (c).}
    \label{fig:eigsplots}
\end{figure*}

In this work, we are mostly focusing on primary bifurcations, \emph{i.e.}, bifurcations arising from the trivial state $(\u_0, p_0, T_0, \B_0, E_0)$ defined by \cref{eq:trivialsol}, and depicted in \cref{fig:bifurcaton_diagram}. These bifurcations occur at critical values, $\Ra_c$, of the bifurcation parameter $\Ra$, which can be computed by re-organizing the eigenvalue problem associated with the trivial state at eigenvalue $\lambda=0$ as a generalized eigenvalue problem:
\begin{align*}
    \begin{bmatrix}
        2 \nabla \cdot \varepsilon(\ ) - Q\, \mathbf{\hat{z}} \times (\cdot \times \mathbf{\hat{z}}) & -\nabla & 0        & 0                     & - Q\,\mathbf{\hat{z}} \times \vphantom{\tilde{E}} \\
        \nabla \cdot                                                                                 & 0       & 0        & 0                     & 0 \vphantom{\tilde{E}}                            \\
        \mathbf{\hat{z}} \cdot                                                                       & 0       & \nabla^2 & 0                     & 0 \vphantom{\tilde{E}}                            \\
        0                                                                                            & 0       & 0        & \nabla (\nabla \cdot) & - \vcurl \vphantom{\tilde{E}}                     \\
        - \cdot \times \mathbf{\hat{z}}                                                              & 0       & 0        & -\scurl               & -I \vphantom{\tilde{E}}
    \end{bmatrix}
     &
    \begin{bmatrix}
        \tilde{\u} \\ \tilde{p} \\  \tilde{T} \\ \tilde{\B}\\ \tilde{E}
    \end{bmatrix} \\
    = \Ra_c
    \begin{bmatrix}
        0 & 0 & - \hat{\mathbf{z}} & 0 & 0 \vphantom{\tilde{E}} \\
        0 & 0 & 0                  & 0 & 0 \vphantom{\tilde{E}} \\
        0 & 0 & 0                  & 0 & 0 \vphantom{\tilde{E}} \\
        0 & 0 & 0                  & 0 & 0 \vphantom{\tilde{E}} \\
        0 & 0 & 0                  & 0 & 0 \vphantom{\tilde{E}}
    \end{bmatrix}
     &
    \begin{bmatrix}
        \tilde{\u} \\ \tilde{p} \\  \tilde{T} \\ \tilde{\B}\\ \tilde{E}
    \end{bmatrix},
\end{align*}
where we interpret $\Ra_c$ as an eigenvalue. Recall that $\u_0=\E_0=0$ and hence we dropped these terms above. We then solve this problem to obtain the first ten critical Rayleigh numbers at $Q=10^3$ and associated eigenmodes (see \cref{fig:eigsplots}(e)). The first critical Chandrasekhar numbers, $Q_c$, can also be computed at $\Ra=10^5$ by solving an analogous eigenvalue problem:
\begin{align*}
    \begin{bmatrix}
        2 \nabla \cdot \varepsilon(\ )  & -\nabla & \Ra\, \hat{\mathbf{z}} & 0                    & 0 \vphantom{\tilde{E}}        \\
        \nabla \cdot                    & 0       & 0                      & 0                    & 0 \vphantom{\tilde{E}}        \\
        \mathbf{\hat{z}} \cdot          & 0       & \nabla^2               & 0                    & 0 \vphantom{\tilde{E}}        \\
        0                               & 0       & 0                      & \nabla(\nabla \cdot) & - \vcurl \vphantom{\tilde{E}} \\
        - \cdot \times \mathbf{\hat{z}} & 0       & 0                      & -\scurl              & -I \vphantom{\tilde{E}}
    \end{bmatrix}
     &
    \begin{bmatrix}
        \tilde{\u} \\ \tilde{p} \\  \tilde{T} \\ \tilde{\B}\\ \tilde{E}
    \end{bmatrix} \\
    = Q_c
    \begin{bmatrix}
        \mathbf{\hat{z}} \times (\cdot \times \mathbf{\hat{z}}) & 0 & 0 & 0 & \mathbf{\hat{z}} \times \vphantom{\tilde{E}} \\
        0                                                       & 0 & 0 & 0 & 0 \vphantom{\tilde{E}}                       \\
        0                                                       & 0 & 0 & 0 & 0 \vphantom{\tilde{E}}                       \\
        0                                                       & 0 & 0 & 0 & 0 \vphantom{\tilde{E}}                       \\
        0                                                       & 0 & 0 & 0 & 0 \vphantom{\tilde{E}}
    \end{bmatrix}
     &
    \begin{bmatrix}
        \tilde{\u} \\ \tilde{p} \\  \tilde{T} \\ \tilde{\B}\\ \tilde{E}
    \end{bmatrix}.
\end{align*}

\section{Results} \label{sec:results}

In this section, we report the bifurcation diagrams and solutions to the magnetic Rayleigh--B\'enard problem discovered by deflation and analyze their behavior as the Rayleigh and Chandrasekhar numbers vary.

\subsection{Location of primary instabilities}

We begin our study by computing the stability of the conducting state~\eqref{eq:trivialsol} to localize the primary instabilities and resulting bifurcations. To do so, we solve the associated eigenvalue problem over a range of Rayleigh and Chandrasekhar numbers, namely $0\leq\Ra\leq 10^5$ at $Q=1$, $0\leq Q\leq 10^3$ at $\Ra=10^5$, and $0\leq\Ra\leq 10^5$ at $Q=10^3$ using the numerical method described in \cref{app:unstableEigenmodes}. The positive real parts of the eigenvalues for the different intervals are reported in panels (a-c) of \cref{fig:eigsplots}, respectively.

We observe in \cref{fig:eigsplots}(a) that eleven supercritical pitchfork bifurcations emanate from the conducting state at $Q=1$ when the Rayleigh number increases to $\Ra=10^5$. This plot is nearly identical to the one computed for the standard Rayleigh B\'enard problem at $Q=0$~\cite[Fig.~1]{Boulle2022} indicating that the effect of the magnetic field on the bifurcation structures is negligible when $Q=1$. In our computation, the eleventh supercritical bifurcation takes place at $\Ra_c^{(11)}\approx 9.95\times 10^4$, while it starts slightly above $\Ra=10^5$ in the standard Rayleigh--B\'enard problem and is therefore not included in~\cite{Boulle2022}.

\begin{figure*}[htbp]
    \centering
    \begin{overpic}[width=\textwidth]{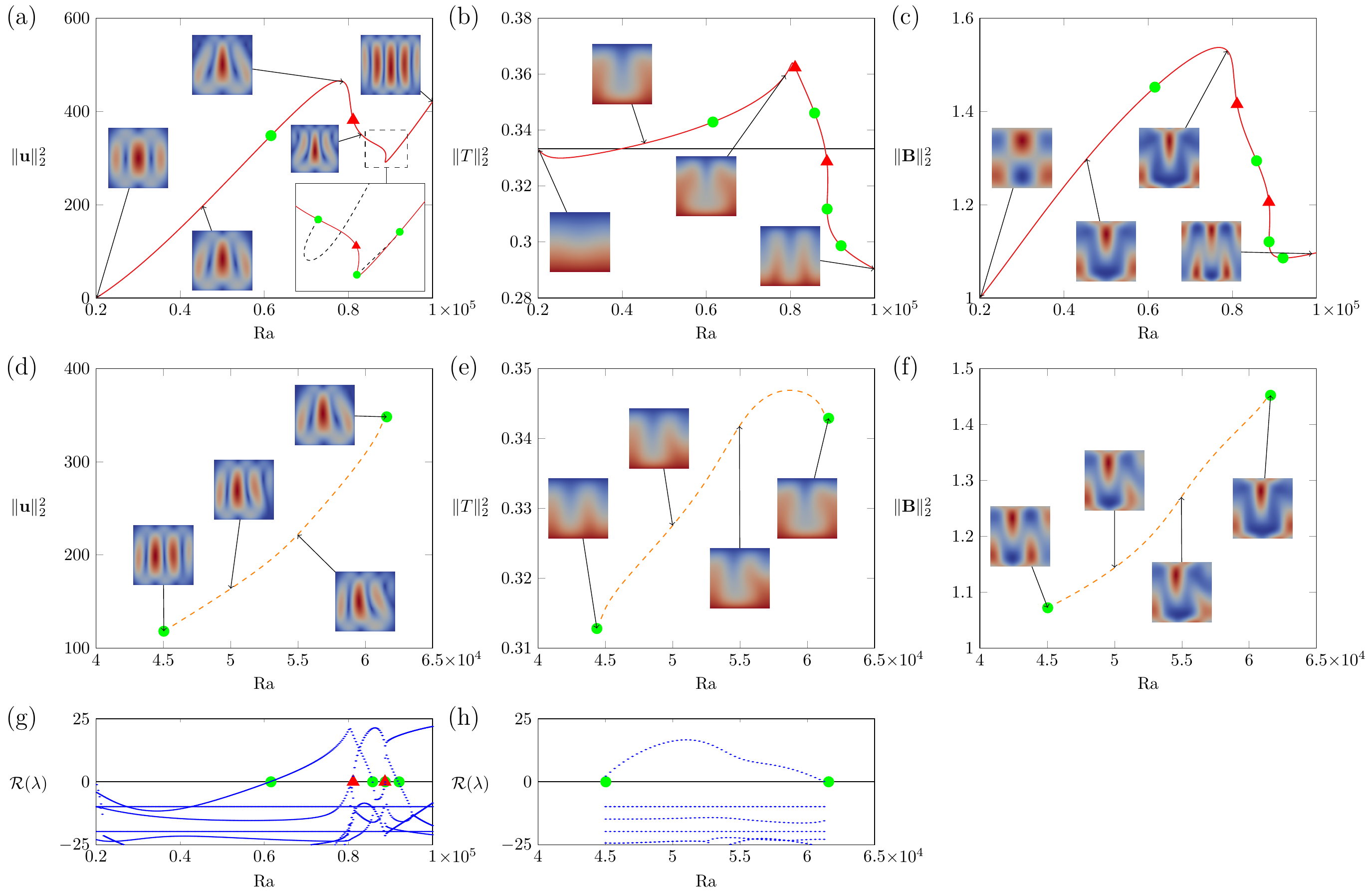}
    \end{overpic}
    \caption{Evolution of the steady states in the first primary branch (1) with respect to the Rayleigh number, where the Chandrasekhar is fixed to $Q=10^3$, illustrated via the kinetic energy (a), potential energy (b), and magnetic energy (c). The plots in these panels display the magnitude of the velocity, temperature, and magnitude of the magnetic field. The color schemes are defined as follows. For the velocity and magnetic fields, blue corresponds to zero while red stands for high magnitude (the color maps are rescaled to the data range). For the temperature, blue corresponds to $T=0$ and red to $T=1$. The corresponding largest growth rates are reported in (g). (d-f) Same as (a-c) but the evolution is for the steady states in the secondary branch (6), with the largest growth rates plotted in (h). The green dots and red triangles indicate the presence of pitchfork bifurcations (eigenvalue $\lambda=0$) and Hopf bifurcations (pair of complex conjugate purely imaginary eigenvalues).}
    \label{fig:bifurcation_Ra_S1000_diagram_branch_3}
\end{figure*}

\cref{fig:eigsplots}(d) displays the velocity magnitude of the first ten eigenmodes corresponding to the first primary bifurcations of the conducting state at $Q=1$, as well as the associated critical Rayleigh numbers (located where the conducting state has a zero eigenvalue). The critical Rayleigh numbers differ only by one percent when compared with the values computed at $Q=0$ and the eigenmodes look visually similar to~\cite[Fig.~2]{Boulle2022}. These critical Rayleigh numbers were obtained following the procedure described in \cref{app:unstableEigenmodes}. The eigenmodes are classified following the number of rolls, $(m_x,m_z)$, of their velocity profile in the horizontal and vertical directions. As an example, the first eigenmode is a $(1,1)$-eigenmode, while the third one is a $(3,1)$-eigenmode since the velocity field has three rolls in the horizontal direction and one in the vertical direction.

We now analyze the effect of the strength of the magnetic field on the primary bifurcations by fixing $\Ra=10^5$ and increasing the Chandrasekhar number to $Q=10^3$ in \cref{fig:eigsplots}(b). We find that the number of unstable eigenmodes decreases to five in this range, leading to the observation that, contrary to the Rayleigh number, increasing the Chandrasekhar number penalizes the birth of primary bifurcations as the growth rates of the conducting state decrease in \cref{fig:eigsplots}(b). The delayed onset of instabilities when increasing $Q$ aligns with the findings of~\cite{Busse1982Stability,Basak2014}, while observed in a different parameter regime.

\cref{fig:eigsplots}(c) displays the positive growth rates of the conducting state for $0\leq \Ra \leq 10^5$  at  $Q=10^3$. The first five primary bifurcations occur at the critical Rayleigh numbers at which there is a zero eigenvalue, \emph{i.e.}, $\Ra_c^{(1)}\approx 2.0\times 10^4$, $\Ra_c^{(2)}\approx 2.3\times 10^4$, $\Ra_c^{(3)}\approx 3.2\times 10^4$, $\Ra_c^{(4)}\approx 5.9\times 10^4$, and $\Ra_c^{(5)}\approx 9.9\times 10^4$. The values of the critical Rayleigh numbers agree with the ones obtained by solving a generalized eigenvalue problem (cf.~\cref{app:unstableEigenmodes}), and the locations of the bifurcations from the conducting state in \cref{fig:bifurcaton_diagram}(c).

\begin{figure*}[htbp]
    \centering
    \begin{overpic}[width=\textwidth]{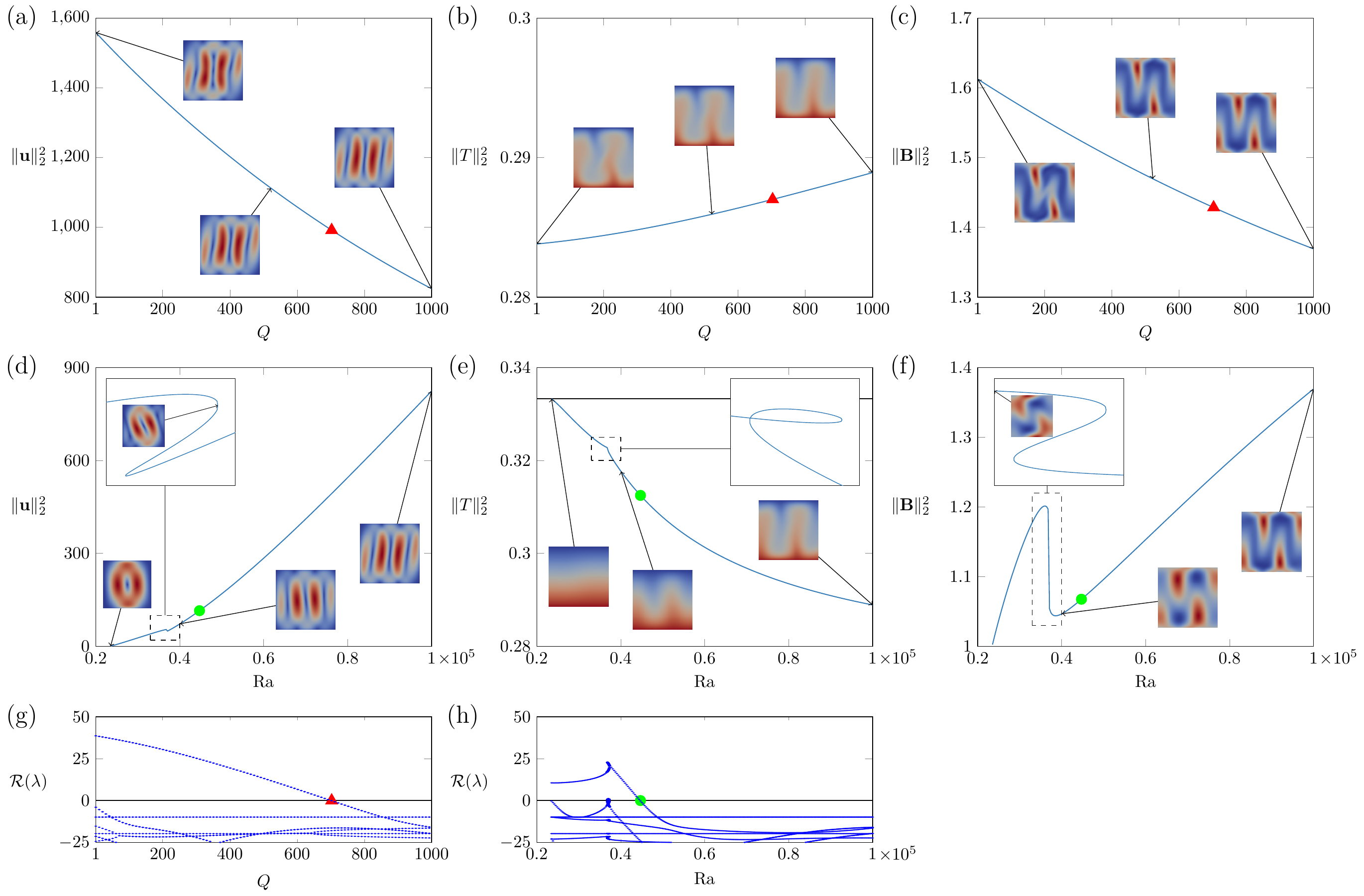}
    \end{overpic}
    \caption{Evolution of the steady states in the second primary branch (2) as $Q$ increases at $\Ra=10^5$, illustrated via the kinetic energy (a), potential energy (b), and magnetic energy (c). (d-f) Same as (a-c) but the evolution is displayed over $\Ra$ at $Q=10^3$. Panel (g) shows the largest growth rates (real parts of the eigenvalues) corresponding to the states in (a-c), while (h) reports the largest growth rates over $\Ra$ at $Q=10^3$, \emph{i.e.}, states in (d-f).}
    \label{fig:branch_1_S}
\end{figure*}

Then, the velocity magnitude of the first five eigenmodes associated with the critical Rayleigh numbers over $0\leq \Ra \leq 10^5$ at $Q=10^3$ is displayed in \cref{fig:eigsplots}(e). It is interesting to observe that the velocity profile (in particular the vortices) of the eigenmodes feature a pattern that is oriented in the (upward) direction of the background magnetic field $\B_0=\hat{\mathbf{z}}$. This indicates that for a larger Chandrasekhar number at $Q = 10^3$, the instabilities aligned with the magnetic field occur at smaller Rayleigh numbers.  In general, the order of eigenmodes changes with growing $Q$. For instance, the $(4,1)$-eigenmode is the seventh eigenmode in \cref{fig:eigsplots}(a) at $Q=1$ but corresponds to the fourth eigenmode at $Q=10^3$. Similarly, we find that the $(1,1)$ and $(2,1)$-eigenmodes are swapped when increasing the Chandrasekhar number from $Q=1$ to $Q=10^3$ (cf.~\cref{fig:eigsplots}(a,d,e)).

We display the bifurcation diagrams of the magnetic Rayleigh--B\'enard problem in \cref{fig:bifurcaton_diagram} for the kinetic, potential, and magnetic energy. In the bifurcation diagrams, we do not report symmetries of solutions generated by the mirror and Boussinesq symmetries. Additionally, since the effect of the magnetic field at $Q=1$ is minor, we just track the evolution of the branches connected to the branches at $Q=10^3$ in \cref{fig:bifurcaton_diagram}(a). We then refer to \cite{Boulle2022} for a detailed bifurcation analysis of the non-magnetic case.

\begin{figure*}[htbp]
    \centering
    \begin{overpic}[width=\textwidth]{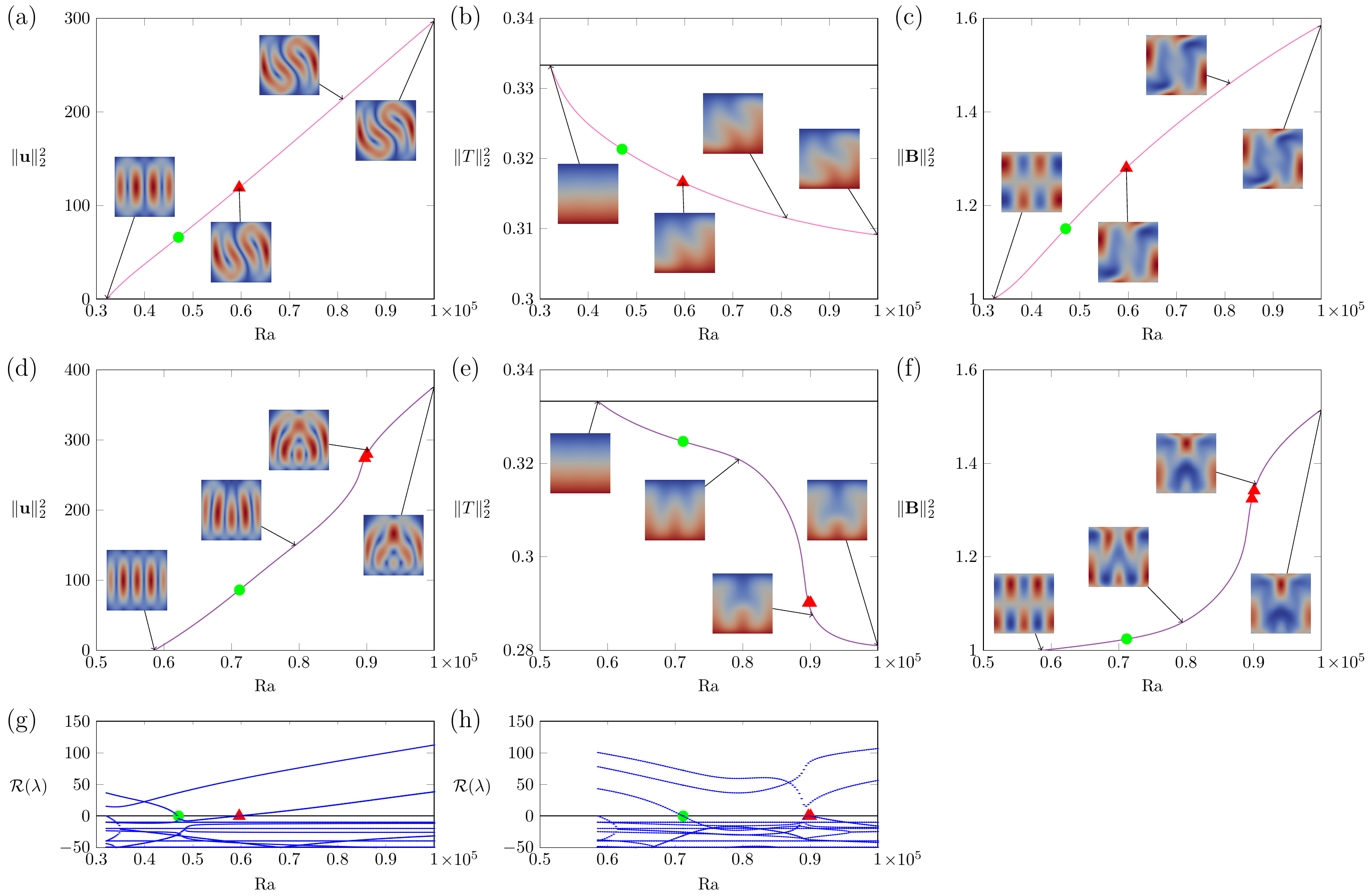}
    \end{overpic}
    \caption{Evolution of the steady states in the third primary branch (3) with respect to the Rayleigh number, where the Chandrasekhar number is fixed to $Q=10^3$, illustrated via the kinetic energy (a), potential energy (b), and magnetic energy (c). The corresponding largest growth rates are reported in (g). (d-f) Same as (a-c) but the evolution is for the steady states in the fourth primary branch (4), with the largest growth rates plotted in (h). The secondary branch (7) emanates at the two bifurcation points and connects the two primary branches.}
    \label{fig:bifurcation_Ra_S1000_diagram_branch_4}
\end{figure*}

\subsection{First and second primary branches}

We begin our study of the solutions computed via deflation by analyzing the evolution of the steady states in branch (1) at Rayleigh number $\Ra=10^5$, illustrated in \cref{fig:bifurcation_Ra_S1000_diagram_branch_3}(a-c). In panel (g), we report the largest growth rates (real part of eigenvalues) associated with the solutions and highlight the location of simple bifurcations in green and Hopf bifurcations by a red dot in the different diagrams. As the Chandrasekhar number increases to $Q=10^3$ in \cref{fig:bifurcation_Ra_S1000_diagram_branch_3}(a-c), the deflation algorithm found a rather surprising evolution of the states in this branch. Hence, while the branch emanates from the $(2,1)$-eigenmode at $\Ra\approx 2\times 10^4$ and $Q=10^3$ by breaking the Boussinesq symmetry but keeping the mirror symmetry, with two convection rolls on the magnitude velocity field, the states in the branch evolve to follow a pattern with four rolls for Rayleigh number larger than $\Ra\approx 8.4\times 10^4$. The latter profile near $\Ra\approx 10^5$ in \cref{fig:bifurcation_Ra_S1000_diagram_branch_3}(a) is reminiscent to the $(4,1)$-eigenmode depicted in \cref{fig:eigsplots}(e). We have investigated this branch by refining the grid resolution and the continuation step size but could not discover additional branches connecting to the fourth eigenmode. To deepen the understanding of the evolution of the states in the branch, we have provided a more detailed diagram in the inlet plot of \cref{fig:bifurcation_Ra_S1000_diagram_branch_3}(a) for the magnitude of the velocity field of branch (1) in the range $8.4\times 10^4\leq \Ra\leq 9.4\times 10^4$. This additional diagram indicates that the correct primary branch was discovered and reported and that there is no jump to secondary branches. There exist two saddle-node bifurcations located at $\Ra \approx 8.87\times 10^4$ and $\Ra \approx 8.89\times 10^4$, leading to an S-shaped curve. Moreover, three pitchfork bifurcations take place in this region at respectively $\Ra\approx 8.47\times 10^4$, $\Ra\approx 8.89\times 10^4$, and $\Ra\approx 9.22\times 10^4$, highlighted by green dots in \cref{fig:bifurcation_Ra_S1000_diagram_branch_3}(g). We have indicated these secondary branches in dashed black curves in \cref{fig:bifurcation_Ra_S1000_diagram_branch_3}(a) to emphasize that the branches have been discovered by deflation but will not be analyzed further in this work. Finally, we note the presence of a Hopf bifurcation at $\Ra \approx 8.9\times 10^4$ and a subcritical pitchfork bifurcation at $\Ra\approx 6.2\times 10^4$. The secondary branch (6) arising from this bifurcation is depicted in \cref{fig:bifurcation_Ra_S1000_diagram_branch_3}(d-f) and will be analyzed later. Based on the plot of the growth rates in \cref{fig:bifurcation_Ra_S1000_diagram_branch_3}(g), we see that this branch is stable for Rayleigh numbers smaller than $6.2\times 10^4$ and believe that all the bifurcations from branch (1) have been obtained. Hence, the bifurcation diagram is complete. Then, as we decrease the Chandrasekhar number towards $Q=1$, we find in \cref{fig:bifurcaton_diagram}(b) that branch (1) undergoes two saddle-node bifurcations around $Q\approx 610$ and $Q\approx 415$. Finally, after fixing $Q=1$ and decreasing the Rayleigh number, we observe in the bifurcation diagrams depicted in \cref{fig:bifurcaton_diagram}(a) that the branch originates from the $(4,1)$-eigenmode at the seventh critical Rayleigh number $\Ra\approx 4.7\times 10^4$, as expected by the profile of the velocity magnitude for $\Ra\geq 9\times 10^4$ in \cref{fig:bifurcation_Ra_S1000_diagram_branch_3}(a).

\begin{figure*}[htbp]
    \centering
    \begin{overpic}[width=\textwidth]{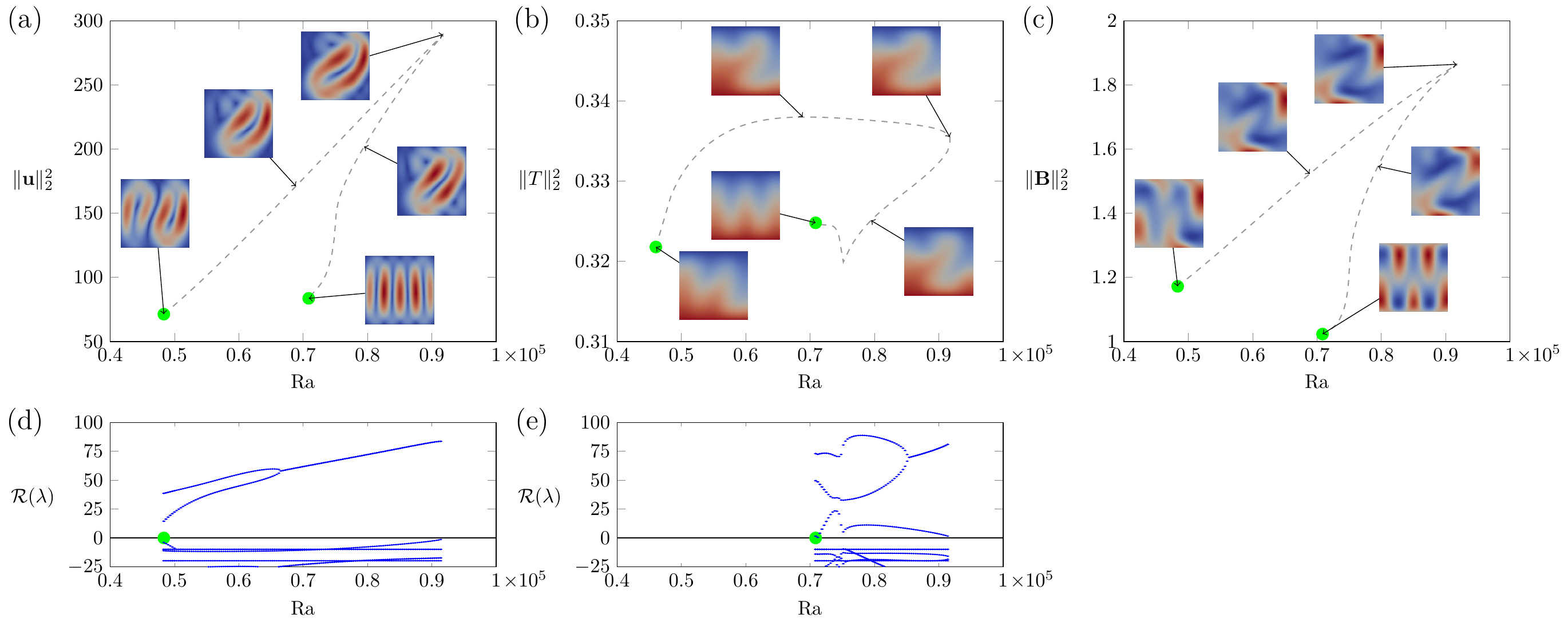}
    \end{overpic}
    \caption{Evolution of the steady states in the secondary branch (7), bifurcating from branches (3) and (4), with respect to the Rayleigh number, where the Chandrasekhar is fixed to $Q=10^3$, and illustrated via the kinetic energy (a), potential energy (b), and magnetic energy (c). This branch features a fold bifurcation at $\Ra\approx 9.17\times 10^4$, with an upper branch in the range $4.83\times 10^4\leq\Ra\leq 9.17\times 10^4$ and a lower branch in the range $7.09\times10^4\leq \Ra\leq 9.17\times 10^4$. The largest growth rates for the upper and lower branches are respectively reported in (d) and (e).}
    \label{fig:bifurcation_Ra_S1000_diagram_branch_8}
\end{figure*}

We proceed with the secondary branch (6), which emanates from branch (1) at $\Ra\approx 6.2\times 10^4$ and connects to branch (2) at $\Ra\approx 4.47\times 10^4$. As we observe in \cref{fig:bifurcation_Ra_S1000_diagram_branch_3}(d-f), this branch has lost the mirror symmetries of its parent branches and evolves from a velocity field with three convection rolls to two convection rolls. Additionally, the branch is weakly unstable over its range of existence with a maximum growth rate of $\mathcal{R}(\lambda)\approx 17$, as confirmed by \cref{fig:bifurcation_Ra_S1000_diagram_branch_3}(h).

We conclude our analysis of the first and second primary branches and their secondary bifurcations by focusing on branch (2) in \cref{fig:branch_1_S}(a-c) and (g). We find that the states in branch (2) originally emanate from the $(3,1)$-eigenmode around $\Ra=0$ and $Q=1$ in \cref{fig:bifurcaton_diagram}(a) and have the Boussinesq symmetry. The profile of the velocity, temperature, and magnetic fields does not change much in \cref{fig:branch_1_S}(a-c) when increasing the Chandrasekhar number as the three convection rolls of the magnitude velocity field, \emph{i.e.}, two smaller rolls on the sides and a stronger in the middle, are preserved from $Q=1$ to $Q=10^3$. Interestingly, we observe in \cref{fig:branch_1_S}(g) that this branch goes through a Hopf bifurcation at $Q\approx 700$ and becomes stable afterward. This indicates that increasing the strength of the magnetic field may help stabilize solutions to the magnetic Rayleigh--B\'enard problem.

Additionally, we ran a direct numerical simulation (DNS) for three states in \cref{fig:branch_1_S}(a-c) at $Q\approx 500$, $Q\approx 700$ (at the Hopf bifurcation point) and $Q\approx1000$ to follow time-dependent states in the unstable, periodic and stable regime. Movies of these simulations are available in the Supplementary Material. To start the DNS, we chose $\epsilon = 0.1$ and perturbed the steady states by $\epsilon$-times the normalized eigenfunctions that correspond to the eigenvalues with the largest real part. At the Hopf point at $Q\approx 700$, the imaginary part of the eigenvalue $\lambda$ with vanishing real part is given by $\mathcal{I}(\lambda)=111.01$. Hence, we expect a time-periodic solution of the length $2\pi/\mathcal{I}(\lambda)\approx 0.0566$, which is verified in the attached movie in the  Supplementary Material. Moreover, for the unstable regime at $Q\approx 500$ and stable regime at $Q\approx 1000$ we observe the expected behavior that the perturbed solution either diverges or converges to the steady state. We discretized in time with the Crank-Nicolson scheme and used a step-size of $\Delta t = 0.01$ for $Q\approx 500$ and $Q\approx1000$. For $Q \approx 700$, we chose  $\Delta t = 0.00566$ such that 10 timesteps correspond to one full period and stop after four full periods.

We then focus on \cref{fig:branch_1_S}(d-f) to study the behavior of the steady states in the branch at $Q=10^3$ as the Rayleigh number decreases to find the linear limit from which the branch arises. One of the eigenvalues (depicted by a green dot in \cref{fig:branch_1_S}(h)) crosses zero as $\Ra\approx 4.47\times 10^4$, which gives birth to a subcritical pitchfork bifurcation, leading to branch (6). Surprisingly, branch (2) undergoes a sequence of two saddle-node bifurcations afterward around $\Ra\approx 3.7\times 10^4$, as confirmed by the inlet plots of \cref{fig:branch_1_S}(d-f) and finally originate from the $(1,1)$ eigenmode by breaking the initial mirror symmetry of the eigenmode. This behavior is counterintuitive as one would have expected branch (2) to bifurcate from the $(3,1)$-eigenmode instead.

\subsection{Third and fourth primary branches}

This section focuses on the third and fourth primary branches arising from the conducting states at $Q=10^3$ over $\Ra\in[0,10^5]$. First, the third primary branch (3) plotted in \cref{fig:bifurcation_Ra_S1000_diagram_branch_4}(a-c) emanates from the $(3,1)$-eigenmode of the conducting state through a supercritical pitchfork bifurcation at the critical Rayleigh number $\Ra\approx 3.2\times 10^4$ (cf.~\cref{fig:eigsplots}(e)). As we see in the magnitude velocity profile of \cref{fig:bifurcation_Ra_S1000_diagram_branch_4}(a), the branch has lost the mirror symmetry of the third eigenmode but kept the Boussinesq symmetry. As the Rayleigh number increases, we observe that the velocity field evolves to feature two convection rolls located symmetrically at the bottom left and top right corners of the domain. Interestingly, this behavior is similar to the evolution of the states in the branch arising from the third eigenmode in the non-magnetic case (see \cite[Fig.~7]{Boulle2022}) and does not seem to be influenced by the presence of an external magnetic field. Then, looking at the growth rates reported in \cref{fig:bifurcation_Ra_S1000_diagram_branch_4}(g), we note the presence of a supercritical pitchfork bifurcation at $\Ra\approx 4.7\times 10^4$, leading to the secondary branch (7) illustrated in \cref{fig:bifurcation_Ra_S1000_diagram_branch_8}, and a Hopf bifurcation located at $\Ra\approx 5.96\times 10^4$. After reaching Rayleigh number $\Ra=10^5$, we study the influence of the magnetic field on steady states in this branch by decreasing the Chandrasekhar number in \cref{fig:bifurcaton_diagram}(b). We observe that the branch bifurcates from branch (5) at $Q\approx 567$ after going through a saddle-node bifurcation around $Q\approx 540$.

Here, branch (5), depicted in green in the bifurcation diagrams of \cref{fig:bifurcaton_diagram}(a,b) is the branch bifurcating from the $(2,2)$-eigenmode at $\Ra_c^{(4)}\approx 2.3\times 10^4$ and $Q=1$, \emph{i.e.}, the magnitude velocity profiles of the steady states feature four symmetric convection rolls, with the mirror and Boussinesq symmetries preserved throughout the branch. As the strength of the magnetic field increases at $\Ra=10^5$, the convection rolls become thinner in the $x$ direction and longer in the buoyancy direction (see~\cite[Fig.~4.7(b)]{LaakmannPhD}). However, we find that this branch dies around $Q\approx 965$ and is therefore not present in the bifurcation diagrams at $Q=10^3$ depicted in \cref{fig:bifurcaton_diagram}(c).

\begin{figure*}[htbp]
    \centering
    \begin{overpic}[width=\textwidth]{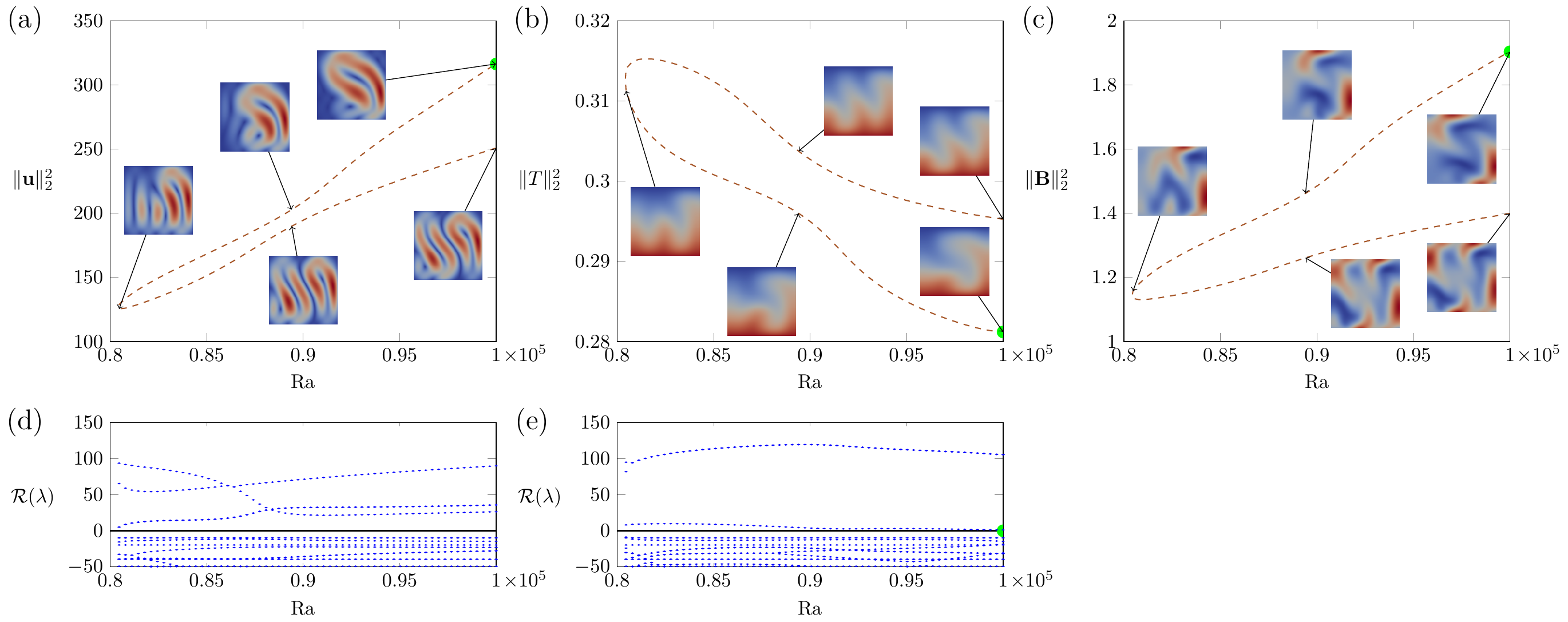}
    \end{overpic}
    \caption{Evolution of the steady states in the disconnected branch (8) with respect to the Rayleigh number, where the Chandrasekhar is fixed to $Q=10^3$, illustrated via the kinetic energy (a), potential energy (b), and magnetic energy (c). This branch features a fold bifurcation at $\Ra\approx 8.04\times 10^4$, with an upper branch in (a) featuring a simple bifurcation at $\Ra\approx 10^5$, as indicated by the green dot. The largest growth rates for the upper and lower branches are respectively reported in (d) and (e).}
    \label{fig:bifurcation_Ra_S1000_diagram_branch_6}
\end{figure*}

Moving to the fourth primary branch (4) illustrated in \cref{fig:bifurcation_Ra_S1000_diagram_branch_4}(d-f), we see that the states in this branch emanate from the $(4,1)$-eigenmode of the conducting state through a supercritical pitchfork bifurcation at $\Ra\approx 5.9\times 10^4$ and $Q=10^3$. Contrary to branch (3), we observe in the magnitude velocity plots of \cref{fig:bifurcation_Ra_S1000_diagram_branch_4}(d) that the Boussinesq symmetry of the eigenmode is broken at the bifurcation but the mirror symmetry is preserved. Despite the birth of the state from the $(4,1)$-eigenmode of the conducting state, the velocity and temperature fields of the states in the branch rapidly evolve to a pattern reminiscent of the one displayed in~\cite[Fig.~13(a,b)]{Boulle2022}. As the Chandrasekhar number decreases in \cref{fig:bifurcaton_diagram}, branch (4) bifurcates from branch (5) through a subcritical pitchfork bifurcation at $Q\approx 330$ after a saddle-node bifurcation at $Q\approx 270$, similarly to branch (3). The connection between branch (4) and branch (5), which itself emanates from the $(2,2)$-eigenmode at $Q=1$, might explain the similarities between the velocity and temperature fields in branch (4) and the branch depicted in~\cite[Fig.~13(a,b)]{Boulle2022}. Hence, the latter branch also bifurcates from a primary branch emanating from the $(2,2)$-eigenmode. Moving back to the study of branch (4) at $Q=10^3$, we observe in \cref{fig:bifurcation_Ra_S1000_diagram_branch_4}(h) that the branch is unstable throughout its range of existence (in terms of the Rayleigh number) but gives rise to the secondary branch (7) at $\Ra\approx 7.1\times 10^4$ through a subcritical pitchfork bifurcation and undergoes two Hopf bifurcations later on at $\Ra\approx 8.97\times 10^4$ and $\Ra\approx 9\times 10^4$.

We now analyze the secondary branch (7) depicted in \cref{fig:bifurcation_Ra_S1000_diagram_branch_8}, which bifurcates from branch (3) at $\Ra\approx 4.7\times 10^4$ through a supercritical pitchfork bifurcation and branch (4) at $\Ra\approx 7.1\times 10^4$ through a subcritical bifurcation. This branch breaks the symmetry of its parent branches and features a large central convection roll with two diagonally opposite smaller rolls in the magnitude velocity field illustrated in \cref{fig:bifurcation_Ra_S1000_diagram_branch_8}, similarly to the states in~\cite[Fig.~14]{Boulle2022} for the non-magnetic case. While the branch is unstable, it goes through a saddle-node bifurcation at $\Ra\approx 9.2\times 10^4$.

\subsection{Disconnected branch}

We conclude the study of the branches of solutions to the magnetic Rayleigh--B\'enard problem at $Q=10^3$ and $\Ra\in[0,10^5]$ with branch (8), illustrated in \cref{fig:bifurcation_Ra_S1000_diagram_branch_6}. This branch is highly unstable, non-symmetric, and has a saddle-node bifurcation at $\Ra\approx 8\times 10^4$. Observing \cref{fig:bifurcation_Ra_S1000_diagram_branch_6}(a), we see that the magnitude velocity profile of the lower branch features two large convection rolls and a smaller roll in the bottom left corner while the upper branch has one dominant convection roll and a smaller one located in the lower part of the domain, similarly to the branch reported in~\cite[Fig.~6]{Boulle2022}. We emphasize that branch (8) is a disconnected branch in the bifurcation diagram for the Rayleigh number at $Q=10^3$ and would be difficult to compute with standard bifurcation tools. This branch was obtained without the use of branch-switching techniques by following our strategy to continue branches at low Chandrasekhar numbers first in \cref{fig:bifurcaton_diagram}(b).

We first analyze the effect of the strength of the magnetic field on the upper part of this branch (in \cref{fig:bifurcation_Ra_S1000_diagram_branch_6}(a)) by tracking the steady states as the Chandrasekhar number decreases to $Q=1$ in \cref{fig:bifurcaton_diagram}(b). We find that the upper part of branch (8) persists throughout the range of Chandrasekhar numbers considered by going through two saddle-node bifurcations at $Q\approx 700$ and $Q\approx 310$. Then, after fixing the Chandrasekhar number to $Q=1$ and decreasing the Rayleigh number, we observe in \cref{fig:bifurcaton_diagram}(a) that the branch finally emanates from branch (5) through a supercritical pitchfork bifurcation at $\Ra\approx 3.4\times 10^3$.

Focusing on the evolution of the lower part of branch (8) as the Rayleigh number is fixed to $\Ra=10^5$ and the Chandrasekhar number decreases in \cref{fig:bifurcaton_diagram}(b),
we observe the presence of two saddle-node bifurcations at $Q\approx 610$ and $Q\approx 310$. Finally, and similarly to branch (3), the lower part of branch (8) bifurcates from branch (5) around $Q\approx 567$. However, the discretization size employed in this work does not allow us to characterize the nature of these bifurcations; either one transcritical bifurcation giving birth to the two branches or two distinct pitchfork bifurcations located very closely.

\section{Conclusions} \label{sec_conclusions}

In this work, we performed a bifurcation analysis of a two-dimensional magnetic Rayleigh--B\'enard problem. By combining deflation with a continuation of steady states at low Chandrasekhar number Q, we were able to compute the entire bifurcation diagram at $Q=10^3$ up to Rayleigh number $\Ra = 10^5$. The profusion of steady states obtained reveals the rich dynamics of the problem, similar to the classical Rayleigh--B{\'e}nard problem, with the occurrence of several pitchfork, Hopf, and saddle-node bifurcations.

Additionally, our numerical scheme allowed us to study the influence of the strength of the magnetic field on the branches and the profile of their velocity, temperature, and magnetic fields by continuing and relating them to steady states at low Chandrasekhar numbers. Our findings show that the presence of a vertical magnetic field tends to privilege primary branches emanating from eigenmodes with convection rolls aligned with the direction of the magnetic field. This leads to a swapping of the eigenmodes of the conducting states as the Chandrasekhar number increases.

Moreover, one of the states computed by deflation and depicted in \cref{fig:branch_1_S} becomes stable as the Chandrasekhar number increases beyond $Q\approx 700$. This demonstrates that increasing the strength of the magnetic field may help stabilize solutions. We also conjecture that one could select certain solution profiles by modifying the direction of the background magnetic field.

A potential extension of this work would be to investigate the dependence of the steady states to the magnetic Rayleigh--B\'enard problem on the magnetic Prandtl number in a parameter regime close to the ones of liquid metal experiments~\cite{burr_muller_2002}. One could also exploit parameter-robust preconditioners~\cite[Sec.~4.3]{LaakmannPhD} to build fast numerical solvers and perform bifurcation analysis at large values of the Chandrasekhar number on finer meshes or in three dimensions. Finally, an interesting direction would be to control the patterns of the magnetic fields towards a desired configuration using PDE-constrained optimization techniques~\cite{boulle2022control,boulle2023optimization}.

\section*{Acknowledgement}
We would like to thank Patrick Farrell for many useful suggestions and discussions. We are grateful to the anonymous reviewers for their constructive comments, which helped us improve the manuscript. This work was supported by an INI-Simons Postdoctoral Research Fellowship and the Office of Naval Research (ONR) under grant N00014-23-1-2729 (N.B.).

\bibliographystyle{elsarticle-num}
\bibliography{biblio}

\providecommand{\noopsort}[1]{}
\begin{thebibliography}{10}
\expandafter\ifx\csname url\endcsname\relax
  \def\url#1{\texttt{#1}}\fi
\expandafter\ifx\csname urlprefix\endcsname\relax\def\urlprefix{URL }\fi
\expandafter\ifx\csname href\endcsname\relax
  \def\href#1#2{#2} \def\path#1{#1}\fi

\bibitem{rayleigh1916}
{\noopsort{Rayleigh}}{Lord Rayleigh},
  \href{https://doi.org/10.1080/14786441608635602}{{On convection currents in a
  horizontal layer of fluid, when the higher temperature is on the under
  side}}, Phil. Mag. S. 32~(192) (1916) 529--546.
\newline\urlprefix\url{https://doi.org/10.1080/14786441608635602}

\bibitem{benard1900}
H.~B{\'e}nard, {Etude exp{\'e}rimentale du mouvement des liquides propageant de
  la chaleur par convection. R{\'e}gime permanent: tourbillons cellulaires}, C.
  r. hebd. s\'eances Acad. sci. Paris 130 (1900) 1004--1007.

\bibitem{benard1927}
H.~B{\'e}nard, {Sur les tourbillons cellulaires et la th{\'e}orie de Rayleigh},
  C. r. hebd. s\'eances Acad. sci. Paris 185 (1927) 1109--1111.

\bibitem{chandrasekhar1961hydrodynamic}
S.~Chandrasekhar, Hydrodynamic and hydromagnetic stability, Oxford University
  Press, 1961.

\bibitem{crosshohenberg93}
M.~C. Cross, P.~C. Hohenberg,
  \href{https://link.aps.org/doi/10.1103/RevModPhys.65.851}{Pattern formation
  outside of equilibrium}, Rev. Mod. Phys. 65 (1993) 851--1112.
\newline\urlprefix\url{https://link.aps.org/doi/10.1103/RevModPhys.65.851}

\bibitem{bodenschatzetal00}
E.~Bodenschatz, W.~Pesch, G.~Ahlers,
  \href{https://doi.org/10.1146/annurev.fluid.32.1.709}{{Recent developments in
  Rayleigh-B{\'e}nard convection}}, Annu. Rev. Fluid Mech. 32~(1) (2000)
  709--778.
\newline\urlprefix\url{https://doi.org/10.1146/annurev.fluid.32.1.709}

\bibitem{ouertatani2008numerical}
N.~Ouertatani, N.~B. Cheikh, B.~B. Beya, T.~Lili,
  \href{https://doi.org/10.1016/j.crme.2008.02.004}{{Numerical simulation of
  two-dimensional Rayleigh--B{\'e}nard convection in an enclosure}}, C. R. Mec.
  336~(5) (2008) 464--470.
\newline\urlprefix\url{https://doi.org/10.1016/j.crme.2008.02.004}

\bibitem{zienicke1998bifurcations}
E.~Zienicke, N.~Seehafer, F.~Feudel,
  \href{https://link.aps.org/doi/10.1103/PhysRevE.57.428}{{Bifurcations in
  two-dimensional Rayleigh-B{\'e}nard convection}}, Phys. Rev. E 57~(1) (1998)
  428.
\newline\urlprefix\url{https://link.aps.org/doi/10.1103/PhysRevE.57.428}

\bibitem{paul2012bifurcation}
S.~Paul, M.~K. Verma, P.~Wahi, S.~K. Reddy, K.~Kumar,
  \href{https://doi.org/10.1142/S0218127412300182}{{Bifurcation analysis of the
  flow patterns in two-dimensional Rayleigh--B{\'e}nard convection}}, Int. J.
  Bifurcat. Chaos 22~(05) (2012) 1230018.
\newline\urlprefix\url{https://doi.org/10.1142/S0218127412300182}

\bibitem{mishra2010patterns}
P.~K. Mishra, P.~Wahi, M.~K. Verma,
  \href{https://doi.org/10.1209%2F0295-5075%2F89%2F44003}{{Patterns and
  bifurcations in low--Prandtl-number Rayleigh-B{\'e}nard convection}}, EPL
  89~(4) (2010) 44003.
\newline\urlprefix\url{https://doi.org/10.1209%2F0295-5075%2F89%2F44003}

\bibitem{ma2006multiplicity}
D.-J. Ma, D.-J. Sun, X.-Y. Yin,
  \href{https://link.aps.org/doi/10.1103/PhysRevE.74.037302}{{Multiplicity of
  steady states in cylindrical Rayleigh-B{\'e}nard convection}}, Phys. Rev. E
  74~(3) (2006) 037302.
\newline\urlprefix\url{https://link.aps.org/doi/10.1103/PhysRevE.74.037302}

\bibitem{boronska2010extreme}
K.~Boro{\'n}ska, L.~S. Tuckerman,
  \href{https://link.aps.org/doi/10.1103/PhysRevE.81.036320}{{Extreme
  multiplicity in cylindrical Rayleigh-B{\'e}nard convection. I. Time
  dependence and oscillations}}, Phys. Rev. E 81~(3) (2010) 036320.
\newline\urlprefix\url{https://link.aps.org/doi/10.1103/PhysRevE.81.036320}

\bibitem{boronska2010extreme2}
K.~Boro{\'n}ska, L.~S. Tuckerman,
  \href{https://link.aps.org/doi/10.1103/PhysRevE.81.036321}{{Extreme
  multiplicity in cylindrical Rayleigh-B{\'e}nard convection. II. Bifurcation
  diagram and symmetry classification}}, Phys. Rev. E 81~(3) (2010) 036321.
\newline\urlprefix\url{https://link.aps.org/doi/10.1103/PhysRevE.81.036321}

\bibitem{puigjaner2004stability}
D.~Puigjaner, J.~Herrero, F.~Giralt, C.~Sim{\'o},
  \href{https://doi.org/10.1063/1.1778031}{Stability analysis of the flow in a
  cubical cavity heated from below}, Phys. Fluids 16~(10) (2004) 3639--3655.
\newline\urlprefix\url{https://doi.org/10.1063/1.1778031}

\bibitem{puigjaner2006bifurcation}
D.~Puigjaner, J.~Herrero, F.~Giralt, C.~Sim{\'o},
  \href{https://link.aps.org/doi/10.1103/PhysRevE.73.046304}{{Bifurcation
  analysis of multiple steady flow patterns for Rayleigh-B{\'e}nard convection
  in a cubical cavity at $Pr= 130$}}, Phys. Rev. E 73~(4) (2006) 046304.
\newline\urlprefix\url{https://link.aps.org/doi/10.1103/PhysRevE.73.046304}

\bibitem{Peterson2008}
J.~W. Peterson, \href{http://hdl.handle.net/2152/18091}{Parallel adaptive
  finite element methods for problems in natural convection}, Ph.D. thesis,
  University of Texas, Austin (5 2008).
\newline\urlprefix\url{http://hdl.handle.net/2152/18091}

\bibitem{keller1977numerical}
H.~B. Keller, Numerical solution of bifurcation and nonlinear eigenvalue
  problems., in: Applications of Bifurcation Theory, Academic Press, 1977, pp.
  359--384.

\bibitem{doedel1981auto}
E.~J. Doedel, {AUTO: A program for the automatic bifurcation analysis of
  autonomous systems}, Congr. Numer 30~(265-284) (1981) 25--93.

\bibitem{uecker2014pde2path}
H.~Uecker, D.~Wetzel, J.~D.~M. Rademacher,
  \href{https://doi.org/10.1017/S1004897900000295}{{pde2path-A Matlab package
  for continuation and bifurcation in 2D elliptic systems}}, Numer.
  Math.-Theory Me. 7~(1) (2014) 58--106.
\newline\urlprefix\url{https://doi.org/10.1017/S1004897900000295}

\bibitem{dijkstra2014numerical}
H.~A. Dijkstra, F.~W. Wubs, A.~K. Cliffe, E.~Doedel, I.~F. Dragomirescu,
  B.~Eckhardt, A.~Y. Gelfgat, A.~L. Hazel, V.~Lucarini, A.~G. Salinger, et~al.,
  \href{https://doi.org/10.4208/cicp.240912.180613a}{{Numerical Bifurcation
  Methods and their Application to Fluid Dynamics: Analysis beyond
  Simulation}}, Commun. Comput. Phys. 15~(1) (2014) 1--45.
\newline\urlprefix\url{https://doi.org/10.4208/cicp.240912.180613a}

\bibitem{tuckerman2000bifurcation}
L.~S. Tuckerman, D.~Barkley, Bifurcation analysis for timesteppers, in:
  Numerical methods for bifurcation problems and large-scale dynamical systems,
  Springer, 2000, pp. 453--466.
\newblock \href {https://doi.org/10.1007/978-1-4612-1208-9_20}
  {\path{doi:10.1007/978-1-4612-1208-9_20}}.

\bibitem{mamun1995asymmetry}
C.~K. Mamun, L.~S. Tuckerman,
  \href{https://doi.org/10.1063/1.868730}{{Asymmetry and Hopf bifurcation in
  spherical Couette flow}}, Phys. Fluids 7~(1) (1995) 80--91.
\newline\urlprefix\url{https://doi.org/10.1063/1.868730}

\bibitem{Deflation2015}
P.~E. Farrell, A.~Birkisson, S.~W. Funke, Deflation techniques for finding
  distinct solutions of nonlinear partial differential equations, SIAM J. Sci.
  Comput. 37~(4) (2015) A2026--A2045.
\newblock \href {https://doi.org/10.1137/140984798}
  {\path{doi:10.1137/140984798}}.

\bibitem{Boulle2022}
N.~Boull\'e, V.~Dallas, P.~E. Farrell, Bifurcation analysis of two-dimensional
  {R}ayleigh-{B}\'enard convection using deflation, Phys. Rev. E 105 (2022)
  055106.
\newblock \href {https://doi.org/10.1103/PhysRevE.105.055106}
  {\path{doi:10.1103/PhysRevE.105.055106}}.

\bibitem{weiss_proctor_2014}
N.~O. Weiss, M.~R.~E. Proctor, Magnetoconvection, Cambridge University Press,
  2014.
\newblock \href {https://doi.org/10.1017/CBO9780511667459}
  {\path{doi:10.1017/CBO9780511667459}}.

\bibitem{Busse1982Stability}
F.~H. Busse, R.~M. Clever, Stability of convection rolls in the presence of a
  vertical magnetic field, Phys. Fluids 25~(6) (1982) 931--935.
\newblock \href {https://doi.org/10.1063/1.863845}
  {\path{doi:10.1063/1.863845}}.

\bibitem{PhysRevLett.52.1774}
S.~Fauve, C.~Laroche, A.~Libchaber, B.~Perrin, Chaotic phases and magnetic
  order in a convective fluid, Phys. Rev. Lett. 52 (1984) 1774--1777.
\newblock \href {https://doi.org/10.1103/PhysRevLett.52.1774}
  {\path{doi:10.1103/PhysRevLett.52.1774}}.

\bibitem{aurnou_olson_2001}
J.~M. Aurnou, P.~L. Olson, {Experiments on Rayleigh--B\'enard convection,
  magnetoconvection and rotating magnetoconvection in liquid gallium}, J. Fluid
  Mech. 430 (2001) 283--307.
\newblock \href {https://doi.org/10.1017/S0022112000002950}
  {\path{doi:10.1017/S0022112000002950}}.

\bibitem{burr_muller_2002}
U.~Burr, U.~M{\"u}ller, {R}ayleigh--{B}énard convection in liquid metal layers
  under the influence of a horizontal magnetic field, J. Fluid Mech. 453 (2002)
  345--369.
\newblock \href {https://doi.org/10.1017/S002211200100698X}
  {\path{doi:10.1017/S002211200100698X}}.

\bibitem{yanagisawa2011spontaneous}
T.~Yanagisawa, Y.~Yamagishi, Y.~Hamano, Y.~Tasaka, Y.~Takeda, {Spontaneous flow
  reversals in Rayleigh-B{\'e}nard convection of a liquid metal}, Phys. Rev. E
  83~(3) (2011) 036307.
\newblock \href {https://doi.org/10.1103/PhysRevE.83.036307}
  {\path{doi:10.1103/PhysRevE.83.036307}}.

\bibitem{Proctor_1982}
M.~R.~E. Proctor, N.~O. Weiss, Magnetoconvection, Rep. Prog. Phys. 45~(11)
  (1982) 1317.
\newblock \href {https://doi.org/10.1088/0034-4885/45/11/003}
  {\path{doi:10.1088/0034-4885/45/11/003}}.

\bibitem{glatzmaier1999role}
G.~A. Glatzmaier, R.~S. Coe, L.~Hongre, P.~H. Roberts, The role of the
  {E}arth's mantle in controlling the frequency of geomagnetic reversals,
  Nature 401~(6756) (1999) 885--890.
\newblock \href {https://doi.org/10.1038/44776} {\path{doi:10.1038/44776}}.

\bibitem{Cattaneo_2003}
F.~Cattaneo, T.~Emonet, N.~Weiss, On the interaction between convection and
  magnetic fields, Astrophys. J. 588~(2) (2003) 1183.
\newblock \href {https://doi.org/10.1086/374313} {\path{doi:10.1086/374313}}.

\bibitem{rucklidge2006mean}
A.~M. Rucklidge, M.~R.~E. Proctor, J.~Prat, Mean flow instabilities of
  two-dimensional convection in strong magnetic fields, Geophys. Astrophys.
  Fluid Dyn. 100~(2) (2006) 121--137.
\newblock \href {https://doi.org/10.1080/03091920600565595}
  {\path{doi:10.1080/03091920600565595}}.

\bibitem{Yang2021}
J.~C. Yang, T.~Vogt, S.~Eckert,
  \href{https://link.aps.org/doi/10.1103/PhysRevFluids.6.023502}{Transition
  from steady to oscillating convection rolls in {R}ayleigh-{B}\'enard
  convection under the influence of a horizontal magnetic field}, Phys. Rev.
  Fluids 6 (2021) 023502.
\newblock \href {https://doi.org/10.1103/PhysRevFluids.6.023502}
  {\path{doi:10.1103/PhysRevFluids.6.023502}}.
\newline\urlprefix\url{https://link.aps.org/doi/10.1103/PhysRevFluids.6.023502}

\bibitem{HAN2018370}
D.~Han, M.~Hernandez, Q.~Wang, Dynamical transitions of a low-dimensional model
  for {R}ayleigh--{B}énard convection under a vertical magnetic field, Chaos
  Solitons Fractals 114 (2018) 370--380.
\newblock \href {https://doi.org/https://doi.org/10.1016/j.chaos.2018.06.027}
  {\path{doi:https://doi.org/10.1016/j.chaos.2018.06.027}}.

\bibitem{NAFFOUTI2014714}
A.~Naffouti, B.~Ben-Beya, T.~Lili, Three-dimensional {R}ayleigh--{B}énard
  magnetoconvection: {E}ffect of the direction of the magnetic field on heat
  transfer and flow patterns, C. R. Mécanique 342~(12) (2014) 714--725.
\newblock \href {https://doi.org/https://doi.org/10.1016/j.crme.2014.09.001}
  {\path{doi:https://doi.org/10.1016/j.crme.2014.09.001}}.

\bibitem{Akhmedagaev2020}
R.~Akhmedagaev, O.~Zikanov, D.~Krasnov, J.~Schumacher, Turbulent
  {R}ayleigh--{B}énard convection in a strong vertical magnetic field, J.
  Fluid Mech. 895 (2020) R4.
\newblock \href {https://doi.org/10.1017/jfm.2020.336}
  {\path{doi:10.1017/jfm.2020.336}}.

\bibitem{Tasaka2016}
Y.~Tasaka, K.~Igaki, T.~Yanagisawa, T.~Vogt, T.~Zuerner, S.~Eckert,
  \href{https://link.aps.org/doi/10.1103/PhysRevE.93.043109}{{Regular flow
  reversals in Rayleigh-B\'enard convection in a horizontal magnetic field}},
  Phys. Rev. E 93 (2016) 043109.
\newblock \href {https://doi.org/10.1103/PhysRevE.93.043109}
  {\path{doi:10.1103/PhysRevE.93.043109}}.
\newline\urlprefix\url{https://link.aps.org/doi/10.1103/PhysRevE.93.043109}

\bibitem{Cioni2000}
S.~Cioni, S.~Chaumat, J.~Sommeria,
  \href{https://link.aps.org/doi/10.1103/PhysRevE.62.R4520}{Effect of a
  vertical magnetic field on turbulent rayleigh-b\'enard convection}, Phys.
  Rev. E 62 (2000) R4520--R4523.
\newblock \href {https://doi.org/10.1103/PhysRevE.62.R4520}
  {\path{doi:10.1103/PhysRevE.62.R4520}}.
\newline\urlprefix\url{https://link.aps.org/doi/10.1103/PhysRevE.62.R4520}

\bibitem{Nandukumar_2015}
Y.~Nandukumar, P.~Pal,
  \href{https://doi.org/10.1209/0295-5075/112/24003}{Oscillatory instability
  and routes to chaos in {R}ayleigh-{B}{\'{e}}nard convection: Effect of
  external magnetic field}, EPL 112~(2) (2015) 24003.
\newblock \href {https://doi.org/10.1209/0295-5075/112/24003}
  {\path{doi:10.1209/0295-5075/112/24003}}.
\newline\urlprefix\url{https://doi.org/10.1209/0295-5075/112/24003}

\bibitem{DAS2019}
S.~Das, K.~Kumar,
  \href{https://www.sciencedirect.com/science/article/pii/S001793101931052X}{Thermal
  flux in unsteady rayleigh-bénard magnetoconvection}, Int. J. Heat Mass
  Transf. 142 (2019) 118413.
\newblock \href
  {https://doi.org/https://doi.org/10.1016/j.ijheatmasstransfer.2019.07.063}
  {\path{doi:https://doi.org/10.1016/j.ijheatmasstransfer.2019.07.063}}.
\newline\urlprefix\url{https://www.sciencedirect.com/science/article/pii/S001793101931052X}

\bibitem{Basak2014}
A.~Basak, R.~Raveendran, K.~Kumar,
  \href{https://link.aps.org/doi/10.1103/PhysRevE.90.033002}{Rayleigh-b\'enard
  convection with uniform vertical magnetic field}, Phys. Rev. E 90 (2014)
  033002.
\newblock \href {https://doi.org/10.1103/PhysRevE.90.033002}
  {\path{doi:10.1103/PhysRevE.90.033002}}.
\newline\urlprefix\url{https://link.aps.org/doi/10.1103/PhysRevE.90.033002}

\bibitem{Nakagawa1957Experiments}
Y.~Nakagawa, S.~Chandrasekhar, Experiments on the inhibition of thermal
  convection by a magnetic field, Proc. R. Soc. London, Ser. A 240~(1220)
  (1957) 108--113.
\newblock \href {https://doi.org/10.1098/rspa.1957.0070}
  {\path{doi:10.1098/rspa.1957.0070}}.

\bibitem{Nakagawa1959Experiments}
Y.~Nakagawa, S.~Chandrasekhar, {Experiments on the instability of a layer of
  mercury heated from below and subject to the simultaneous action of a
  magnetic field and rotation. II}, Proc. R. Soc. London, Ser. A 249~(1256)
  (1959) 138--145.
\newblock \href {https://doi.org/10.1098/rspa.1959.0012}
  {\path{doi:10.1098/rspa.1959.0012}}.

\bibitem{knobloch_weiss_costa_1981}
E.~Knobloch, N.~O. Weiss, L.~N.~D. Costa, Oscillatory and steady convection in
  a magnetic field, J. Fluid Mech. 113 (1981) 153--186.
\newblock \href {https://doi.org/10.1017/S0022112081003443}
  {\path{doi:10.1017/S0022112081003443}}.

\bibitem{clever1989nonlinear}
R.~M. Clever, F.~H. Busse, Nonlinear oscillatory convection in the presence of
  a vertical magnetic field, J. Fluid Mech. 201 (1989) 507--523.
\newblock \href {https://doi.org/10.1017/S0022112089001023}
  {\path{doi:10.1017/S0022112089001023}}.

\bibitem{boussinesq1903theorie}
J.~Boussinesq, \href{https://books.google.co.uk/books?id=O08NmgEACAAJ}{Theorie
  analytique de la Chaleur}, Gauthier-Villars, 1903.
\newline\urlprefix\url{https://books.google.co.uk/books?id=O08NmgEACAAJ}

\bibitem{Oberbeck1879}
A.~Oberbeck, \href{https://doi.org/10.1002/andp.18792430606}{Ueber die
  {W}\"{a}rmeleitung der {F}l\"{u}ssigkeiten bei {B}er\"{u}cksichtigung der
  {S}tr\"{o}mungen infolge von {T}emperaturdifferenzen}, Ann. Phys. 243~(6)
  (1879) 271--292.
\newblock \href {https://doi.org/10.1002/andp.18792430606}
  {\path{doi:10.1002/andp.18792430606}}.
\newline\urlprefix\url{https://doi.org/10.1002/andp.18792430606}

\bibitem{Hu2020}
K.~Hu, W.~Qiu, K.~Shi, Convergence of a {B}-{E} based finite element method for
  {MHD} models on {L}ipschitz domains, J. Comput. Appl. Math. 368 (2020)
  112477.
\newblock \href {https://doi.org/10.1016/j.cam.2019.112477}
  {\path{doi:10.1016/j.cam.2019.112477}}.

\bibitem{Brackbill1980}
J.~Brackbill, D.~Barnes, The effect of nonzero {$\nabla$ $\cdotp \mathrm{B}$}
  on the numerical solution of the magnetohydrodynamic equations, J. Comput.
  Phys. 35~(3) (1980) 426--430.
\newblock \href {https://doi.org/10.1016/0021-9991(80)90079-0}
  {\path{doi:10.1016/0021-9991(80)90079-0}}.

\bibitem{taylor1973numerical}
C.~Taylor, P.~Hood, {A numerical solution of the Navier-Stokes equations using
  the finite element technique}, Comput. Fluids 1~(1) (1973) 73--100.
\newblock \href {https://doi.org/10.1016/0045-7930(73)90027-3}
  {\path{doi:10.1016/0045-7930(73)90027-3}}.

\bibitem{Raviart1977}
P.~A. Raviart, J.~M. Thomas, A mixed finite element method for second order
  elliptic problems, in: Mathematical Aspects of Finite Element Methods,
  Springer Berlin Heidelberg, 1977, pp. 292--315.
\newblock \href {https://doi.org/10.1007/bfb0064470}
  {\path{doi:10.1007/bfb0064470}}.

\bibitem{Firedrake}
F.~Rathgeber, D.~A. Ham, L.~Mitchell, M.~Lange, F.~Luporini, A.~T.~T. Mcrae,
  G.-T. Bercea, G.~R. Markall, P.~H.~J. Kelly, Firedrake: automating the finite
  element method by composing abstractions, ACM Trans. Math. Softw. 43~(3)
  (2016) 1--27.
\newblock \href {https://doi.org/10.1145/2998441} {\path{doi:10.1145/2998441}}.

\bibitem{balay2019}
S.~Balay, S.~Abhyankar, M.~F. Adams, J.~Brown, P.~Brune, K.~Buschelman,
  L.~Dalcin, V.~Eijkhout, W.~D. Gropp, D.~Karpeyev, et~al., {PETS}c users
  manual, Tech. Rep. ANL-95/11 - Revision 3.15, Argonne National Laboratory
  (2021).
\newblock \href {https://doi.org/10.2172/1814627} {\path{doi:10.2172/1814627}}.

\bibitem{robinson2017molecular}
M.~Robinson, C.~Luo, P.~E. Farrell, R.~Erban, A.~Majumdar, From molecular to
  continuum modelling of bistable liquid crystal devices, Liq. Cryst.
  44~(14-15) (2017) 2267--2284.
\newblock \href {https://doi.org/10.1080/02678292.2017.1290284}
  {\path{doi:10.1080/02678292.2017.1290284}}.

\bibitem{emerson2018computing}
D.~B. Emerson, P.~E. Farrell, J.~H. Adler, S.~P. MacLachlan, T.~J. Atherton,
  Computing equilibrium states of cholesteric liquid crystals in elliptical
  channels with deflation algorithms, Liq. Cryst. 45~(3) (2018) 341--350.
\newblock \href {https://doi.org/10.1080/02678292.2017.1365385}
  {\path{doi:10.1080/02678292.2017.1365385}}.

\bibitem{charalampidis2018computing}
E.~Charalampidis, P.~Kevrekidis, P.~E. Farrell, {Computing stationary solutions
  of the two-dimensional Gross--Pitaevskii equation with deflated
  continuation}, Commun. Nonlinear Sci. Numer. Simul. 54 (2018) 482--499.
\newblock \href {https://doi.org/10.1016/j.cnsns.2017.05.024}
  {\path{doi:10.1016/j.cnsns.2017.05.024}}.

\bibitem{charalampidis2020bifurcation}
E.~Charalampidis, N.~Boull{\'e}, P.~E. Farrell, P.~G. Kevrekidis, {Bifurcation
  analysis of stationary solutions of two-dimensional coupled Gross--Pitaevskii
  equations using deflated continuation}, Commun. Nonlinear Sci. Numer. Simul.
  87 (2020) 105255.
\newblock \href {https://doi.org/10.1016/j.cnsns.2020.105255}
  {\path{doi:10.1016/j.cnsns.2020.105255}}.

\bibitem{boulle2020deflation}
N.~Boull{\'e}, E.~G. Charalampidis, P.~E. Farrell, P.~G. Kevrekidis,
  {Deflation-based identification of nonlinear excitations of the
  three-dimensional Gross-Pitaevskii equation}, Phys. Rev. A 102~(5) (2020)
  053307.
\newblock \href {https://doi.org/10.1103/PhysRevA.102.053307}
  {\path{doi:10.1103/PhysRevA.102.053307}}.

\bibitem{boulle2022two}
N.~Boull{\'e}, I.~Newell, P.~E. Farrell, P.~G. Kevrekidis,
  \href{https://doi.org/10.48550/arXiv.2208.05703}{{Two-Component 3D Atomic
  Bose-Einstein Condensates Support Complex Stable Patterns}}, arXiv preprint
  arXiv:2208.05703 (2022).
\newline\urlprefix\url{https://doi.org/10.48550/arXiv.2208.05703}

\bibitem{Stewart2002}
G.~W. Stewart, A {K}rylov--{S}chur algorithm for large eigenproblems, SIAM J.
  Matrix Anal. Appl. 23~(3) (2002) 601--614.
\newblock \href {https://doi.org/10.1137/S0895479800371529}
  {\path{doi:10.1137/S0895479800371529}}.

\bibitem{SLEPc}
V.~Hernandez, J.~E. Roman, V.~Vidal, Slepc: A scalable and flexible toolkit for
  the solution of eigenvalue problems, ACM Trans. Math. Softw. 31~(3) (2005)
  351--362.
\newblock \href {https://doi.org/10.1145/1089014.1089019}
  {\path{doi:10.1145/1089014.1089019}}.

\bibitem{hale2012dynamics}
J.~K. Hale, H.~Ko{\c{c}}ak, {Dynamics and Bifurcations}, Springer New York, NY,
  1991.
\newblock \href {https://doi.org/10.1007/978-1-4612-4426-4}
  {\path{doi:10.1007/978-1-4612-4426-4}}.

\bibitem{LaakmannPhD}
F.~Laakmann, Discretisations and preconditioners for magnetohydrodynamics
  models, Ph.D. thesis, University of Oxford (2022).

\bibitem{boulle2022control}
N.~Boull{\'e}, P.~E. Farrell, A.~Paganini, Control of bifurcation structures
  using shape optimization, SIAM J. Sci. Comput. 44~(1) (2022) A57--A76.

\bibitem{boulle2023optimization}
N.~Boull{\'e}, P.~E. Farrell, M.~E. Rognes, {Optimization of Hopf bifurcation
  points}, SIAM J. Sci. Comput. 45~(3) (2023) B390--B411.

\end{thebibliography}

\end{document}